\documentclass[aps,pra,twocolumn,superscriptaddress,floatfix,10pt]{revtex4-2}

\usepackage{amsmath,amssymb,bm,graphicx,booktabs}
\usepackage[colorlinks=true, linkcolor=blue, citecolor=blue, urlcolor=blue]{hyperref}

\graphicspath{{figures/}}

\newcommand{\Tr}{\operatorname{Tr}}
\newcommand{\dd}{\mathrm d}
\newcommand{\ee}{\mathrm e}
\newcommand{\ii}{\mathrm i}
\newcommand{\ket}[1]{|#1\rangle}
\newcommand{\bra}[1]{\langle #1|}
\newcommand{\CRE}{C_{\rm RE}}
\newcommand{\Clog}{C_{\log l_1}}

\newcommand{\RRE}{R_{\rm RE}}
\newcommand{\Rfr}{R_{\rm fr}}

\begin{document}

\title{Resource-based reliability of quantum-coherent devices}

\author{Bowen Sun}
\affiliation{Institute of Physics, Beijing National Laboratory for Condensed Matter Physics,\\
Chinese Academy of Sciences, Beijing 100190, China}
\affiliation{School of Physical Sciences, University of Chinese Academy of Sciences, Beijing 100049, China}

\author{D. L. Zhou}
\email[]{zhoudl72@iphy.ac.cn}
\affiliation{Institute of Physics, Beijing National Laboratory for Condensed Matter Physics,\\
Chinese Academy of Sciences, Beijing 100190, China}
\affiliation{School of Physical Sciences, University of Chinese Academy of Sciences, Beijing 100049, China}

\date{\today}

\begin{abstract}
Quantum devices can lose the resource needed for a specified function without entering a fixed failure subspace.  We propose a resource-based reliability framework for this situation, quantifying the remaining task-relevant resource relative to a stated ideal capacity.  The resource measure is selected through its operational connection to the task and the allowed operations.  We develop this approach for a trusted quantum random-number generator and a calibrated two-port interferometer.  Relative-entropy coherence determines the optimal asymptotic yield of standard coherent qubits under incoherent operations for the former, while logarithmic $l_1$ coherence quantifies the fringe resource of the latter and determines its calibrated correct-port probability.  An exactly solvable multi-spin spin-boson model illustrates how environmental records affect these resources at the same isolated-spin dephasing strength.  Collective noise preserves coherence within magnetization sectors for product inputs, but enhances the decay of Greenberger--Horne--Zeilinger branch coherence.  Finite-record simulations connect the resource definitions to measurement counts and show that increasing the sample size reduces statistical fluctuations without eliminating bias from an incorrect many-spin noise model.  The resulting framework provides a task-specific route to reliability assessment when a fixed failure projector is insufficient, while keeping retained resource capacity distinct from a classical survival probability.
\end{abstract}

\maketitle

\section{Introduction}

Reliability assessment requires a physical criterion for whether a device can perform its specified function.  Classical reliability expresses this criterion through the probability of survival up to a given time under stated operating conditions~\cite{rausand2020system,zio2009reliability,rocchi2017reliability,bain2017statistical,pham2023springer}.  Projector-defined survival and failure subspaces provide one formulation for quantum devices~\cite{cui2023quantum}.  When the failure subspace is absorbing, its population has a direct failure-probability interpretation.  Our earlier dissipative spin-chain model realizes this situation: amplitude damping makes the fully decayed ground state absorbing, and the probability that at least one excitation remains is a survival probability~\cite{sun2026spinchain}.

A fixed failure subspace, however, need not represent the loss of the resource required by a quantum device.  In a chosen basis, incoherent density operators form a convex resource-free set, not the states supported on one proper Hilbert subspace~\cite{baumgratz2014quantifying,chitambar2019quantum}.  Pure dephasing can reduce coherence while leaving every population in that basis unchanged.  For a random-number source, a balanced raw histogram therefore need not reveal the remaining coherent resource.  For a two-port interferometer, neither output port is intrinsically a failure outcome, even as the interference fringe disappears.  These examples motivate assessing the resource that supports the specified task when occupation of a fixed failure subspace is insufficient.

We propose to define resource reliability by the fraction of an operationally justified resource that remains relative to an ideal reference capacity.  The task, reference basis, allowed operations, and normalization are fixed before assessment.  The conceptual contribution is to use a resource measure as the link between quantum-state degradation and device reliability, without requiring a failure projector.  This is not a new resource monotone or a universal replacement for a survival probability.  Its justification rests on the resource--task connection, and its relation to a classical event probability must be established for the specified use of the device.

We make this construction explicit for two coherence-enabled tasks.  In a trusted quantum random-number generator (QRNG), relative-entropy coherence gives the optimal asymptotic yield of standard coherent qubits under incoherent operations~\cite{winter2016operational}.  Normalizing that yield by the ideal input-register capacity defines the resource reliability of the reference architecture.  A labeled output-slot construction gives its asymptotic probability interpretation, with output quality controlled separately by the distillation error.  In an ideal two-port interferometer, path-basis $l_1$ coherence determines the calibrated visibility~\cite{bera2015}.  Logarithmic $l_1$ coherence provides an additive resource measure~\cite{rana2016logarithmic}, from which the correct-port probability follows at a specified operating point.  The all-correct probability over several uses additionally requires independent complete preparations and bath resets.

The multi-spin spin-boson model supplies an exactly solvable setting for comparing these reliability measures under collective and spatially resolving dephasing.  Matching the isolated-spin attenuation isolates the effect of environmental correlations.  The calculations show why coherence retained by product inputs under collective noise does not imply protection of Greenberger--Horne--Zeilinger (GHZ) branch coherence.  The full-phase readout also separates stored basis coherence from performance under a particular phase encoding.

Finally, we connect resource reliability to finite measurement records.  These nonlinear state functions cannot generally be represented by one fixed single-copy observable, but they can be assessed using suitable measurements and calibrated models.  We give a two-phase counting estimator for the one-spin interferometer and a model-assisted QRNG estimator based on an isolated reference probe and a validated many-spin correlation model.  The QRNG simulations estimate the asymptotic resource benchmark, not the yield of an implemented finite-block distillation protocol.  A model-mismatch test demonstrates that more single-spin data can reduce statistical uncertainty without identifying the many-spin resource.  Thus the assessment procedure and its trust assumptions are essential to the proposed reliability description.

Section~\ref{sec:framework} defines the resource-reliability framework, and Sec.~\ref{sec:spinboson} introduces the dynamics.  Sections~\ref{sec:qrng} and~\ref{sec:interferometer} develop the two device examples and their analytical and numerical results.  Section~\ref{sec:assessment} addresses finite-record assessment.  Section~\ref{sec:conclusion} summarizes the scope of the approach and its possible extensions.

\section{Resource reliability}
\label{sec:framework}

\subsection{From failure subspaces to resource-free sets}

In classical reliability theory, let $T$ denote the random time at which the specified function first fails.  The reliability function is
\begin{equation}
  R(t)=P(T>t),
\end{equation}
so $R(t)$ is the probability that the system has not failed by time $t$.  
For a system initially functioning and with no repair included in the model, $R(0)=1$ and $R(t)$ is nonincreasing. 
 When $R(t)>0$ is differentiable, the hazard rate is
\begin{equation}
  h(t)=-\frac{\dd}{\dd t}\ln R(t).
\end{equation}
It is the instantaneous conditional failure rate: given survival up to $t$, the probability of failure in $[t,t+\dd t)$ is $h(t)\dd t+o(\dd t)$.  
Equivalently, $R(t)=R(0)\exp[-\int_0^t h(s)\dd s]$ whenever the stated differentiability conditions hold.
If a quantum system has an absorbing failure subspace $\mathcal H_F$ with projector $P_F$, a direct quantum analogue is
\begin{equation}
  R_{\rm sub}(t)=1-\Tr[P_F\rho(t)].
  \label{eq:subspace_reliability}
\end{equation}
Our earlier dissipative spin-chain work belongs to this class~\cite{sun2026spinchain}.
The device stores excitation energy, and the all-ground state is the absorbing failure state.  The reliability is the probability that at least one excitation remains.

For coherence-based devices, fix a reference or readout basis $\{\ket{i}\}$.
The incoherent states form the convex set
\begin{equation}
  \mathcal I=
  \left\{\delta=\sum_i p_i \ket{i}\bra{i}\right\}.
  \label{eq:incoherent_set_general}
\end{equation}
This is a resource-free set, not a Hilbert subspace, so the expectation value of a fixed projector $P_F$ generally cannot report the retained amount or loss of coherence. 
Resource monotones are defined relative to specified free operations, and an operational interpretation, when available, depends on a specified task. 
A resource-reliability definition must therefore state the task, the allowed operations, and the chosen monotone~\cite{chitambar2019quantum}.
For example, generalized robustness quantifies the maximal advantage in a specified class of subchannel discrimination tasks~\cite{takagi2019operational}. Such results motivate an operational resource--task connection, but do not by themselves define the reliability of an arbitrary device.

Accordingly, before fitting data, we fix the reference basis, allowed free operations, operational sector $\mathcal S$, performance task, chosen resource monotone, assessment protocol, and device decision rule, such as a throughput or error threshold. 
Together these specifications determine the reported quantity, its normalization, and the decision target, and are held fixed throughout the analysis.

Let $\mathcal Q(\rho)$ be the nonnegative resource monotone chosen for that task.
We restrict attention to trajectories $\rho(t)\in\mathcal S$, where $\mathcal S$ is an operational sector with finite, strictly positive $\mathcal Q_{\max}=\max_{\rho\in\mathcal S}\mathcal Q(\rho)$.
We define the corresponding resource reliability as the retained capacity
\begin{equation}
  R_{\mathcal Q}(t)=
  \frac{\mathcal Q[\rho(t)]}{\mathcal Q_{\max}},
  \label{eq:resource_reliability}
\end{equation}
The ratio is dimensionless, lies in $[0,1]$, and assigns unit value to a sector state attaining $\mathcal Q_{\max}$. It is a normalization convention, not a new monotone. 
The operational meaning of $\mathcal Q$, the sector $\mathcal S$, and the measurement record specify the reported capacity. 
A different task, even on the same Hilbert space, requires a different specification.

Equation~\eqref{eq:resource_reliability} is a capacity fraction, not in general a survival probability $P(T>t)$.
It has a classical event interpretation only when the specified task supplies one, as in the independently repeated two-port task below.
It is a degradation curve when the relevant noise maps are resource non-generating for the chosen task and the observed trajectory is nonincreasing.
For non-Markovian dynamics that returns coherence to the device, the corresponding $R_{\mathcal Q}(t)$ can increase~\cite{chanda2016delineating}.
This is a physically meaningful recovery of capacity, but it is not a classical survival function.

When $R_{\mathcal Q}(t)>0$ is differentiable and nonincreasing on a stated interval, define the resource hazard rate by
\begin{equation}
  h_{\mathcal Q}(t)=
  -\frac{\dd}{\dd t}\ln R_{\mathcal Q}(t).
  \label{eq:resource_hazard}
\end{equation}
We use \emph{resource hazard rate} for the instantaneous relative decay rate of the remaining resource, extending the logarithmic form of the classical hazard rate.  It is nonnegative on the stated nonincreasing intervals and does not require $R_{\mathcal Q}(0)=1$.  A classical failure-hazard interpretation additionally requires a specified failure-time model linking resource reliability to a survival probability.  A probability of task availability at a fixed time does not by itself establish this interpretation.  If resource backflow makes $R_{\mathcal Q}(t)$ increase, the same logarithmic derivative becomes negative and describes resource recovery.

\section{Multi-spin spin-boson dynamics}
\label{sec:spinboson}

We use a longitudinal spin-boson model as the common open-system platform~\cite{leggett1987dynamics,caldeira1983path,weiss2012quantum,breuer2002theory}.  During the phase-accumulation stage of the QRNG and interferometer, the total Hamiltonian is
\begin{equation}
  H=H_S+H_B+V,
  \label{eq:total_hamiltonian}
\end{equation}
where the spin system of $M$ two-level spins is described by
\begin{equation}
  H_S=\frac{\hbar\omega_z}{2}\sum_{i=1}^M\sigma_i^z,
  \label{eq:spin_system_hamiltonian}
\end{equation}
and the bosonic environment by
\begin{equation}
  H_B=\sum_k\hbar\omega_k b_k^\dagger b_k,
  \label{eq:bath_hamiltonian}
\end{equation}
The longitudinal spin--environment interaction is
\begin{equation}
  V=
  \frac{\hbar}{2}\sum_{i=1}^M\sigma_i^z
  \sum_k\left(g_{ik}b_k^\dagger+g_{ik}^*b_k\right).
  \label{eq:spin_boson_coupling}
\end{equation}
Here $\omega_z$ is the spin splitting, $\sigma_i^z$ is the Pauli operator of spin $i$, $b_k^\dagger$ and $b_k$ create and annihilate a boson of frequency $\omega_k$, and $g_{ik}$ is the coupling of spin $i$ to mode $k$.  The product $\sigma_z$ basis is denoted by
\begin{equation}
  \ket{\bm s}=\ket{s_1,\ldots,s_M},\qquad s_i=\pm1.
\end{equation}
For the stated $H_S$, $[H_S,V]=0$, so the dynamics is exactly solvable as a pure-dephasing problem~\cite{palma1996quantum,duan1998reducing,reina2002decoherence}.  The driven preparation and analysis stages of Sec.~\ref{sec:interferometer} use $\sigma_x$ pulses and do not satisfy this commutation condition.
For a factorized initial state with a thermal bath,
\begin{equation}
  \rho(0)=\rho_S(0)\otimes\rho_B,
\end{equation}
the reduced matrix elements are
\begin{equation}
  \rho_{\bm s\bm s'}(t)
  =
  \rho_{\bm s\bm s'}(0)
  \exp[-\ii\Phi_{\bm s\bm s'}(t)]
  \exp[-\Gamma_{\bm s\bm s'}(t)].
  \label{eq:exact_dephasing_solution}
\end{equation}
Here $\Phi_{\bm s\bm s'}$ is the deterministic phase, including free precession and the bath response and $\Gamma_{\bm s\bm s'}\geq0$ is the decoherence exponent.  The latter can be written as
\begin{equation}
  \Gamma_{\bm s\bm s'}(t)=
  \frac14\sum_{ij}
  (s_i-s'_i)\gamma_{ij}(t)(s_j-s'_j),
  \label{eq:general_dephasing_exponent}
\end{equation}
where
\begin{equation}
  \gamma_{ij}(t)=
  \int_0^\infty \dd\omega\,
  J_{ij}(\omega)
  \frac{1-\cos\omega t}{\omega^2}
  \coth\frac{\beta\hbar\omega}{2}.
  \label{eq:gamma_ij}
\end{equation}
The cross spectral density is $J_{ij}(\omega)=\sum_k g_{ik}g_{jk}^*\delta(\omega-\omega_k)$.  Its diagonal entries set the isolated-spin noise spectrum, whereas its off-diagonal entries specify how strongly the bath records different spins jointly.  To compare bath-record structures at the same isolated-spin dephasing strength, we take identical diagonal spectra and use the Ohmic reference spectrum
\begin{equation}
  J_{ii}(\omega)\equiv J(\omega)=\eta\omega\ee^{-\omega/\omega_c},
  \label{eq:ohmic_spectrum}
\end{equation}
where $\eta$ is the dimensionless coupling strength, $\omega_c$ is the cutoff frequency, and $\beta$ in Eq.~\eqref{eq:gamma_ij} is the inverse bath temperature.  Device-specific numerical values are given with the QRNG and interferometer results.

We compare two limiting bath-record structures.  For a common record, every spin couples identically to each bath mode,
\begin{equation}
  g_{ik}=g_k.
  \label{eq:common_coupling}
\end{equation}
Thus $J_{ij}(\omega)=J(\omega)$ for every pair $i,j$.  Substitution into Eq.~\eqref{eq:gamma_ij} then gives the spin-independent dephasing kernel
\begin{equation}
  \gamma_{ij}(t)=\Gamma(t).
  \label{eq:common_bath}
\end{equation}
Here
\begin{equation}
  \Gamma(t)\equiv
  \int_0^\infty \dd\omega\,
  J(\omega)\frac{1-\cos\omega t}{\omega^2}
  \coth\frac{\beta\hbar\omega}{2}.
  \label{eq:common_dephasing_kernel}
\end{equation}
Thus the bath records only the total magnetization.  For spatially resolving records, we model the cross spectral density through the frequency-independent positive spatial kernel
\begin{equation}
  J_{ij}(\omega)=J(\omega)C_{ij}.
  \label{eq:spatial_spectral_factorization}
\end{equation}
It follows that
\begin{equation}
  \gamma_{ij}(t)=\Gamma(t)C_{ij},
  \label{eq:spatial_dephasing_kernel}
\end{equation}
and we choose
\begin{equation}
  C_{ij}=\exp\!\left(-\frac{|i-j|a}{\xi}\right),
  \label{eq:spatial_record_kernel}
\end{equation}
with lattice spacing $a$ and correlation length $\xi$.  The factorization makes $C_{ii}=1$, so every model has the same $\gamma_{ii}(t)=\Gamma(t)$.  The common-bath limit is $C_{ij}=1$ and the short-correlation limit approaches a local record.
These are experimentally meaningful design limits: coupling the spins predominantly to one shared resonator, or to a mode whose phase is nearly constant across the device, approximates a common record.  Separate local modes, or a propagating mode with resolvable phase differences, move the device toward the spatially resolving limit~\cite{palma1996quantum,duan1998reducing,reina2002decoherence}.  Common and independent baths also produce different dissipation and decoherence dynamics in coupled-qubit systems~\cite{wang2015dissipation}, although those systems need not satisfy the pure-dephasing commutation condition used here.  In the present model, the distinction between common and spatially resolving records controls whether many-body environmental records are compressed or extensive.

For the real symmetric spatial kernels used here, write $C=(C_{ij})$.  Completing the square for each oscillator gives the deterministic phase in Eq.~\eqref{eq:exact_dephasing_solution}~\cite{reina2002decoherence},
\begin{equation}
  \begin{aligned}
    \Phi_{\bm s\bm s'}(t)
    &=\frac{\omega_z t}{2}\sum_i(s_i-s_i')\\
    &\quad-\frac{\Theta(t)}{4}
    \left(\bm s^{\mathsf T}C\bm s-\bm s'^{\mathsf T}C\bm s'\right).
  \end{aligned}
  \label{eq:full_dephasing_phase}
\end{equation}
Here $\mathsf T$ denotes transpose, and $\Theta(t)$ is the dimensionless bath-response kernel,
\begin{equation}
  \Theta(t)=\int_0^\infty\dd\omega\,
  J(\omega)\frac{\omega t-\sin(\omega t)}{\omega^2}.
  \label{eq:bath_response_kernel}
\end{equation}
For the Ohmic spectrum in Eq.~\eqref{eq:ohmic_spectrum},
\begin{equation}
  \Theta(t)=\eta\left[\omega_c t-\arctan(\omega_c t)\right].
  \label{eq:ohmic_bath_response}
\end{equation}
Unlike $\Gamma(t)$, this phase kernel is independent of temperature.  These expressions use the Hamiltonian as written, without an additional spin--spin term that compensates the bath-induced interaction.  The first term in Eq.~\eqref{eq:full_dephasing_phase} describes local free precession and leaves all spin entropies unchanged.  The second term contains joint spin phases when $C_{ij}\ne0$ for $i\ne j$.  Although it leaves the full-state spectrum unchanged, it can change the one-spin marginal entropies and must therefore be retained when evaluating interspin correlations.

\section{Quantum random-number generator}
\label{sec:qrng}

\subsection{Device operation and resource reliability}
\label{sec:qrng_definition}

A quantum random-number generator (QRNG) converts the intrinsic randomness of a quantum measurement into a classical bit string.  Ma et al. identify quantum measurement itself as the physical source of this randomness and emphasize that the type of randomness claim depends on the assumed trust in devices, noise, and the information available to a potential adversary~\cite{ma2016quantum}.  Accordingly, we specify both the spin implementation and its trust assumptions before defining reliability.  The readout of each spin is a measurement of the Pauli operator $\sigma_z$ in the basis $\{\ket{0},\ket{1}\}$.  The standard coherent state $\ket{+}=(\ket{0}+\ket{1})/\sqrt2$ gives the two outcomes with equal probability.  A general coherent input can give biased outcomes, whereas a diagonal mixed state can also give a balanced histogram.  Thus a balanced raw-bit distribution alone does not identify the coherence resource available for the task.

\begin{figure*}[t]
  \centering
  \includegraphics[width=0.98\textwidth]{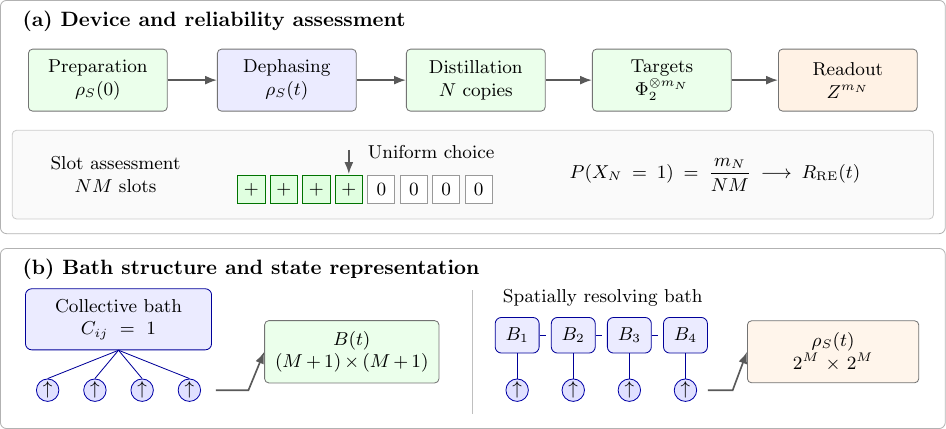}
  \caption{QRNG operation and resource-reliability assessment. (a) A prepared $M$-spin register undergoes pure dephasing. Incoherent processing of $N$ independent copies supplies $m_N$ target coherent qubits for readout. The conversion error is $\epsilon_N$, and the residual state $\tau_N$ is excluded from the yield. The separate assessment strip contains $NM$ logical slots, with $+$ denoting target slots ($\Phi_2$) and $0$ incoherent placeholders ($\ket{0}\bra{0}$). Uniform sampling gives the exact target-label probability $m_N/(NM)$, approaching $\RRE(t)$ for optimal processing with $\epsilon_N\to0$. Slot counts are schematic. (b) For the symmetric product input, a collective bath ($g_{ik}=g_k$, hence $C_{ij}=1$) permits the magnetization-sector representation $B(t)$. Spatially resolving noise generally requires the full state and correlation kernel. The boxes show the respective matrix dimensions. Local bath regions $B_j$ may be correlated, as indicated by dashed links. Panels (a) and (b) correspond to Secs.~\ref{sec:qrng_definition} and~\ref{sec:qrng_model_results}, respectively.}
  \label{fig:qrng_operational}
\end{figure*}

The implementation in Fig.~\ref{fig:qrng_operational}(a) uses an $M$-spin source register, where $M$ is the number of two-level spins in the Hamiltonian of Sec.~\ref{sec:spinboson}.  The states $\rho_S(0)$ and $\rho_S(t)$ describe this entire register at preparation and after a storage time $t$.  They are reduced states because the bath degrees of freedom have been traced out:
\begin{equation}
  \begin{aligned}
    \rho_S(t)&=\mathcal N_t[\rho_S(0)]\\
    &=\Tr_B\!\left\{U(t)[\rho_S(0)\otimes\rho_B]U^\dagger(t)\right\}.
  \end{aligned}
  \label{eq:qrng_reduced_state}
\end{equation}
Here $\Tr_B$ denotes the partial trace over the bath, $U(t)=\exp(-\ii Ht/\hbar)$ is the joint evolution under the total Hamiltonian $H$, and $\rho_B$ is the initial thermal bath state.  The channel $\mathcal N_t$ is the longitudinal pure-dephasing evolution derived in Sec.~\ref{sec:spinboson}.  It preserves the populations in the product $\sigma_z$ basis while attenuating off-diagonal elements.  In particular, pure dephasing can leave the raw measurement probabilities unchanged even when the available coherence decreases.

We assume that preparation, the readout basis, and the relevant noise parameters and correlations have been characterized.  The bath is unobserved during operation, and no bath record is available for feedback or counted as an output resource.  A trusted bath still produces spin--bath correlations and decoherence.  Taking its partial trace is not an assertion that correlations or information outside the spins are absent.  The randomness interpretation below assumes that a potential adversary has no access to the source, bath, or internal processing records.  This specifies an intrinsic-randomness capacity under a trusted-device model.  A claim of private randomness against an adversary with such information requires a separate security analysis~\cite{ma2016quantum,ma2019coherence}. Source-independent QRNG protocols demonstrate certification with an untrusted source~\cite{cao2016sourceindependent}, but that security framework is not assumed here.

To quantify the remaining capacity, we consider a reference architecture that distills standard coherent qubits before readout, consistent with the coherence-as-resource formulation of distillable coherence~\cite{winter2016operational,yuan2015intrinsic}.  At each fixed $M$ and $t$, the input to this processing stage comprises $N$ independently and identically prepared registers.  Experimentally, this assumption requires independent environments or an effective bath reset between preparations.  Correlations among the $M$ spins within each register are allowed.  The processing operations are taken from the class of incoherent operations in the product readout basis: every recorded branch maps a diagonal input to a diagonal output~\cite{baumgratz2014quantifying,winter2016operational}.  No additional coherent state is supplied after preparation.  The resulting capacity is an ideal processing benchmark. Restricted control, imperfect gates, and detector losses must be included separately when evaluating an implemented device.

The operation and resource conversion are summarized by
\begin{equation}
  \begin{aligned}
    \rho_S(0)&\xrightarrow{\mathcal N_t}\rho_S(t),\\
    \rho_S(t)^{\otimes N}&\xrightarrow[\epsilon_N]{\mathcal E_N}
    \Phi_2^{\otimes m_N}\otimes\tau_N,\\
    \Phi_2^{\otimes m_N}&\xrightarrow{\sigma_z\ {\rm readout}}Z^{m_N}.
  \end{aligned}
  \label{eq:qrng_workflow}
\end{equation}
Here the superscript $\otimes N$ denotes $N$ tensor-product copies, $\mathcal E_N$ is an incoherent distillation operation, $\Phi_2=\ket{+}\bra{+}$ is one standard coherent qubit, and $m_N$ is the integer number of target output qubits.  The symbol $\tau_N$ denotes the density operator of a residual output register, which is not counted in the yield and can be discarded or reset to an incoherent state.  The string $Z^{m_N}$ is the classical readout of the target register.  Although the subscript emphasizes the copy number $N$, $m_N$ also depends on the fixed $M$, $t$, source state, protocol, and allowed error.  It is not independent of the source size.

The distillation arrow in Eq.~\eqref{eq:qrng_workflow} denotes approximate conversion within the
distillable-coherence framework~\cite{winter2016operational}, with its error made explicit below.  We use
trace-preserving operations and a fixed target length, without retaining only successful trials.  We consider a sequence of conversions
\begin{equation}
  \mathcal E_N[\rho_S(t)^{\otimes N}]
  \underset{\epsilon_N}{\simeq}\Phi_2^{\otimes m_N}\otimes\tau_N
  \label{eq:qrng_distillation_protocol}
\end{equation}
whose trace-distance error vanishes:
\begin{equation}
  \begin{aligned}
    \epsilon_N={}&\frac12\bigl\|
      \mathcal E_N[\rho_S(t)^{\otimes N}]\\
      &\quad-\Phi_2^{\otimes m_N}\otimes\tau_N
    \bigr\|_1\xrightarrow{N\to\infty}0.
  \end{aligned}
  \label{eq:qrng_distillation_error}
\end{equation}
The norm $\|A\|_1=\Tr\sqrt{A^\dagger A}$ is the trace norm of an operator $A$.  Any CPTP map does not increase trace distance, and both readout and discarding the residual register are CPTP maps.  By monotonicity, the trace-distance error between the actual target readout state and the ideal target remains bounded by $\epsilon_N$ under the stated trust assumptions.  The last line of Eq.~\eqref{eq:qrng_workflow} is the ideal target readout, and the same error bound applies to the actual output.  For classical bit strings, this is equivalent to the usual total-variation distance.

The resource-free states form the set $\mathcal I_z$ of density operators diagonal in the product readout basis.  The relative-entropy coherence is
\begin{equation}
  \CRE(\rho)=S[\Delta(\rho)]-S(\rho),
  \label{eq:cre_def}
\end{equation}
where $S(\rho)=-\Tr(\rho\log_2\rho)$ is the von Neumann entropy in bits, $\Tr$ is the trace, and $\Delta$ removes off-diagonal elements in that basis.  The distillable coherence $C_d(\rho)$ is the optimal asymptotic number of standard coherent qubits per input copy, with vanishing conversion error.  The Winter--Yang theorem gives~\cite{winter2016operational}
\begin{equation}
  C_d(\rho)=\CRE(\rho).
  \label{eq:distillable_coherence}
\end{equation}
For a sequence of protocols approaching this optimal rate at fixed $M$ and $t$,
\begin{equation}
  \lim_{N\to\infty}\frac{m_N}{N}
  =C_d[\rho_S(t)]=\CRE[\rho_S(t)].
  \label{eq:qrng_yield_per_register}
\end{equation}
The denominator counts source registers, each containing $M$ spins.  The average yield need not be an integer because the conversion can act jointly on many registers.  This equality is an optimal asymptotic statement, not an exact finite-$N$ yield or a guarantee for an arbitrary $\mathcal E_N$. At finite $N$ and prescribed error, the achievable output length instead belongs to the one-shot coherence-distillation problem for the entire $N$-copy input~\cite{zhao2019oneshot}.

An ideal register $\ket{+}^{\otimes M}$ supplies $M$ standard coherent qubits.  With this fixed reference capacity, the resource reliability of Sec.~\ref{sec:framework} becomes
\begin{equation}
  \RRE(t)=\frac{\CRE[\rho_S(t)]}{M}.
  \label{eq:qrng_reliability_general}
\end{equation}
It therefore has the operational expression
\begin{equation}
  \lim_{N\to\infty}\frac{m_N}{NM}
  =\frac{C_d[\rho_S(t)]}{M}
  =\frac{\CRE[\rho_S(t)]}{M}
  =\RRE(t).
  \label{eq:qrng_yield_fraction}
\end{equation}
Equation~\eqref{eq:qrng_yield_per_register} counts the output per register, whereas Eq.~\eqref{eq:qrng_yield_fraction} counts it per input spin.  The normalization lies in $[0,1]$, vanishes on $\mathcal I_z$, and equals one for a maximally coherent register.  Its numerator is fixed by an optimal conversion task rather than chosen solely for mathematical convenience.  At each fixed $t$, the longitudinal dephasing channel is incoherent in the readout basis and supplies no additional coherence.  This does not assume that the reduced dynamics is divisible into incoherent channels between arbitrary intermediate times.  Preparation bias can give $\RRE(0)<1$.

To connect this capacity to a classical event, fix the following assessment protocol before comparing devices.  Choose a fixed-length distillation protocol with $0\leq m_N\leq NM$, retain its $m_N$ labeled target slots, and fill the remaining $NM-m_N$ logical slots with incoherent placeholders.  Let $\mathcal A_N(t)$ be the set of target-slot labels and let $J_N$ be a label selected uniformly from $\{1,\ldots,NM\}$.  Define the binary variable
\begin{equation}
  X_N(t)=
  \begin{cases}
    1,&J_N\in\mathcal A_N(t),\\
    0,&J_N\notin\mathcal A_N(t).
  \end{cases}
  \label{eq:qrng_slot_indicator}
\end{equation}
The event $X_N=1$ means that the sampled slot is assigned to the protocol's target output, whose quality is specified by $\epsilon_N$.  Its classical probability is exactly
\begin{equation}
  \begin{aligned}
    P_{\rm usable}^{(N)}(t)
    &=P[X_N(t)=1]=\mathbb E[X_N(t)]\\
    &=\frac{m_N}{NM}
    \xrightarrow[\epsilon_N\to0]{N\to\infty}\RRE(t),
  \end{aligned}
  \label{eq:qrng_usable_probability}
\end{equation}
for the optimal sequence above. $\mathbb E$ denotes classical expectation over the uniform choice.  Thus the resource reliability is the asymptotic probability of supplying a usable coherent-bit slot when one ideal-capacity slot is requested uniformly.  At finite $N$, the exact equality concerns the target label, not a measurement certifying that an individual qubit is exactly $\Phi_2$.  Output quality must still be reported through $\epsilon_N$.  The assessor's random choice is external to the QRNG and contributes no bits to its reported yield.

This construction gives the capacity fraction the same binary-event form used in classical task availability.  The event is specified on the processed output at a given $t$.  It does not define the first failure time of one device or make $\RRE(t)$ a survival law $P(T>t)$.  Nor does it turn $\CRE$ into the expectation value of a fixed single-copy observable.  The probability interpretation depends on the fixed task, normalization, and sampling rule.  A lower-yield implemented protocol supplies its own achievable slot probability.  Reaching Eq.~\eqref{eq:qrng_reliability_general} additionally requires optimal asymptotic processing.  These distinctions preserve the classical event interpretation without identifying raw-bit balance with resource reliability.

\subsection{Analytical expressions and numerical comparisons}
\label{sec:qrng_model_results}

Figure~\ref{fig:qrng_operational}(b) connects the reliability definition to the bath structures of Sec.~\ref{sec:spinboson}.  We first obtain the entropy expression common to these models, then compare collective and spatially resolving records at the same isolated-spin dephasing strength, i.e. with \(\gamma_{ii}(t)=\Gamma(t)\) and therefore \(C_{ii}=1\).  The preparation is
\begin{equation}
  \ket{\psi(q)}=
  \left(\sqrt{1-q}\ket{0}+\sqrt q\ket{1}\right)^{\otimes M},
  \label{eq:biased_product_state}
\end{equation}
with $\rho_S(0)=\ket{\psi(q)}\bra{\psi(q)}$ and $0<q<1$.  Here $q$ is the initial $\ket{1}$ population of each spin.  The function $h_2(q)$ denotes the binary entropy,
\begin{equation}
  h_2(q)=-q\log_2q-(1-q)\log_2(1-q).
  \label{eq:binary_entropy}
\end{equation}
Initially, $S[\rho_S(0)]=0$ and $\CRE[\rho_S(0)]=M h_2(q)$.  Pure dephasing preserves the full product-basis population distribution, so $S[\Delta(\rho_S(t))]=M h_2(q)$ at every time.  Substitution into Eq.~\eqref{eq:qrng_reliability_general} gives
\begin{equation}
  \RRE^{(q)}(t)=h_2(q)-\frac{S[\rho_S(t)]}{M}.
  \label{eq:qrng_reliability}
\end{equation}
The superscript $(q)$ records the preparation bias.  The entire noise dependence is now contained in the entropy of the reduced spin state, which is determined by the dephasing matrix elements in Eq.~\eqref{eq:exact_dephasing_solution}.

This reliability, normalized to the ideal $M$-spin capacity, obeys $\RRE^{(q)}(0)=h_2(q)$ and is used in the numerical comparisons below.  Its slot probability is specified in Sec.~\ref{sec:qrng_definition}.  When $\RRE^{(q)}(t)>0$ is differentiable and nonincreasing, the resource hazard rate of Sec.~\ref{sec:framework} becomes
\begin{equation}
  \begin{aligned}
    h_{\rm RE}^{(q)}(t)
    &=-\frac{\dd}{\dd t}\ln\RRE^{(q)}(t)\\
    &=\frac{\dot S[\rho_S(t)]}{\CRE[\rho_S(t)]}.
  \end{aligned}
  \label{eq:qrng_hazard}
\end{equation}
Here $\dot S[\rho_S(t)]=\dd S[\rho_S(t)]/\dd t$, and the denominator $\CRE[\rho_S(t)]=M\RRE^{(q)}(t)$ is the remaining total coherence capacity.  The resource hazard rate describes the fractional loss of that capacity on the stated interval.  If non-Markovian backflow causes $\RRE^{(q)}(t)$ to increase, then $\dot S[\rho_S(t)]<0$ and the logarithmic derivative is negative, indicating resource recovery.  It cannot then be interpreted as a conventional classical first-failure hazard.

\textit{Collective bath.} In the collective-coupling limit $g_{ik}=g_k$, every spin couples identically to each bath mode and $\gamma_{ij}(t)=\Gamma(t)$ [Eq.~\eqref{eq:common_bath}].  The nonnegative kernel $\Gamma(t)$ is shared by every spin pair.  Sharing one physical bath does not by itself imply collective dephasing.  The condition used here is identical coupling of every spin to every bath mode, under which the bath records only total magnetization.  Let $\bm s$ label a product configuration and let $n_{\bm s}$ be its number of spins in $\ket{1}$, so that $\sum_i s_i=M-2n_{\bm s}$.  Substituting $\gamma_{ij}(t)=\Gamma(t)$ into Eq.~\eqref{eq:general_dephasing_exponent} gives
\begin{equation}
  \begin{aligned}
  \Gamma_{\bm s\bm s'}(t)
  &=\frac{\Gamma(t)}{4}
  \left[\sum_i(s_i-s'_i)\right]^2\\
  &=\Gamma(t)(n_{\bm s}-n_{\bm s'})^2,
  \end{aligned}
  \label{eq:qrng_collective_exponent}
\end{equation}
Thus configurations in the same magnetization sector experience no relative dephasing.  For example, $\ket{01}$ and $\ket{10}$ couple identically to this bath. This preservation is an instance of the decoherence-free-subspace mechanism under collective noise~\cite{duan1998reducing,lidar1998decoherencefree}.

For the input in Eq.~\eqref{eq:biased_product_state}, define the sector population and its normalized coherent vector by
\begin{equation}
  p_n(q)=\binom{M}{n}q^n(1-q)^{M-n},
  \label{eq:qrng_sector_population}
\end{equation}
\begin{equation}
  \ket{u_n}=\binom{M}{n}^{-1/2}
  \sum_{\bm s:n_{\bm s}=n}\ket{\bm s}.
  \label{eq:qrng_sector_vector}
\end{equation}
Here $n=0,\ldots,M$ is the sector label and $\binom{M}{n}$ counts its configurations.  The state remains supported on the $M+1$ vectors $\ket{u_n}$.  After removing the sector-dependent unitary phase factors $\exp[-\ii\Phi_{nn'}(t)]$ inherited from Eq.~\eqref{eq:exact_dephasing_solution}, which leave the spectrum and relative-entropy coherence unchanged, its nonzero spectrum is that of the $(M+1)\times(M+1)$ matrix
\begin{equation}
  B_{nn'}(t)=
  \sqrt{p_n(q)p_{n'}(q)}
  \exp[-\Gamma(t)(n-n')^2].
  \label{eq:compressed_B}
\end{equation}
This is an exact state-representation compression for the present symmetric product input, whose sector weights are binomial and whose sector vectors are uniform.  It is not the distillation operation on $N$ copies in Sec.~\ref{sec:qrng_definition}.  Since $S[\rho_S(t)]=S[B(t)]$, the collective-bath reliability is
\begin{equation}
  R_{{\rm RE},{\rm com}}^{(q)}(t)
  =h_2(q)-\frac{S[B(t)]}{M},
  \label{eq:common_qrng_reliability}
\end{equation}
where the subscript $\mathrm{com}$ denotes the collective bath.

The fixed diagonal of $B$ gives the entropy bound
\begin{equation}
  S[B(t)]\leq H_{\rm binomial}(M,q)\leq\log_2(M+1),
  \label{eq:qrng_sector_entropy_bound}
\end{equation}
where $H_{\rm binomial}(M,q)=-\sum_{n=0}^M p_n(q)\log_2p_n(q)$ is the Shannon entropy of the sector populations.  The second inequality is the maximum-entropy bound for a distribution over $M+1$ outcomes~\cite{cover2006elements}.  For the continuum Ohmic bath used here, arbitrarily low-frequency modes make $\Gamma(t)$ diverge at long times.  Hence $\exp[-\Gamma(t)(n-n')^2]$ vanishes for $n\ne n'$, so different sectors completely dephase and $B(t)$ becomes diagonal.  Thus
\begin{equation}
  \begin{aligned}
    S[B(\infty)]&=H_{\rm binomial}(M,q)\\
    &\simeq\tfrac12\log_2[2\pi e Mq(1-q)].
  \end{aligned}
\end{equation}
This approximation follows from the Gaussian limit of the binomial distribution and assumes fixed $0<q<1$ and $Mq(1-q)\gg1$~\cite{knessl1998entropy}.  For $M>1$, the long-time reliability is
\begin{equation}
  \begin{aligned}
  R_{{\rm RE},{\rm com}}^{(q)}(\infty)
  &=h_2(q)-\frac{H_{\rm binomial}(M,q)}{M}\\
  &\simeq h_2(q)-\frac{1}{2M}\log_2[2\pi eMq(1-q)].
  \end{aligned}
  \label{eq:qrng_compression_scaling}
\end{equation}
Since $B(t)$ approaches the time-independent matrix $B(\infty)$, $\dot S[\rho_S(t)]\to0$, while $\CRE[\rho_S(t)]\to M R_{{\rm RE},{\rm com}}^{(q)}(\infty)>0$.  Thus
\begin{equation}
  h_{\rm RE}^{(q)}(t)\xrightarrow{t\to\infty}0.
  \label{eq:qrng_common_hazard_limit}
\end{equation}
The collective environment resolves only $M+1$ magnetization sectors, leaving coherence within each sector.  Thus its reliability approaches $h_2(q)$ with an $O(\log M/M)$ finite-size correction.  This size limit concerns one register and is distinct from the independent-copy limit $N\to\infty$ used to define distillation.

\textit{Spatially resolving bath.} For the spatially resolving model of Eqs.~\eqref{eq:spatial_spectral_factorization}--\eqref{eq:spatial_record_kernel}, the dephasing kernel is $\gamma_{ij}(t)=\Gamma(t)C_{ij}$.  Thus $\gamma_{ii}(t)=\Gamma(t)$ is the same as in the collective model, whereas the off-diagonal kernels depend on the spin pair.  Configurations with the same total magnetization can therefore leave distinguishable environmental records, so the above sector reduction generally fails.  In the strictly local limit $C_{ij}=\delta_{ij}$, where $\delta_{ij}$ is the Kronecker delta, the kernel becomes $\gamma_{ij}(t)=\Gamma(t)\delta_{ij}$ and the input evolves as a product of identical one-spin reduced states, $\rho_S(t)=\rho_1(t)^{\otimes M}$.  Up to a diagonal phase rotation, $\rho_1(t)$ has off-diagonal magnitude $\sqrt{q(1-q)}\,\ee^{-\Gamma(t)}$ and eigenvalues
\begin{equation}
  \lambda_\pm(t)=\frac{1\pm
  \sqrt{(1-2q)^2+4q(1-q)\ee^{-2\Gamma(t)}}}{2}.
  \label{eq:qrng_local_eigenvalues}
\end{equation}
It follows that
\begin{equation}
  R_{{\rm RE},{\rm loc}}^{(q)}(t)
  =h_2(q)-h_2[\lambda_+(t)],
  \label{eq:qrng_local_reliability}
\end{equation}
where $\mathrm{loc}$ denotes the local bath.  Differentiating Eq.~\eqref{eq:qrng_local_reliability} gives, for $t>0$ and positive reliability,
\begin{equation}
  \begin{aligned}
  h_{{\rm RE},{\rm loc}}^{(q)}(t)
  &=\frac{2q(1-q)\dot\Gamma(t)\ee^{-2\Gamma(t)}}
  {[2\lambda_+(t)-1]R_{{\rm RE},{\rm loc}}^{(q)}(t)}
  \\
  &\quad\times\log_2\frac{\lambda_+(t)}{1-\lambda_+(t)}.
  \end{aligned}
  \label{eq:qrng_local_hazard}
\end{equation}
Both expressions are independent of $M$.  Under complete dephasing, $R_{{\rm RE},{\rm loc}}^{(q)}(t)\to0$ and its leading decay is proportional to $\ee^{-2\Gamma(t)}$, so $h_{{\rm RE},{\rm loc}}^{(q)}(t)/\dot\Gamma(t)\to2$.

For the finite-range kernel $C_{ij}=\exp[-|i-j|a/\xi]$, the state generally does not factorize.  Let $w_{\bm s}=q^{n_{\bm s}}(1-q)^{M-n_{\bm s}}$ be the fixed population of configuration $\bm s$.  Removing the diagonal unitary phases in Eq.~\eqref{eq:exact_dephasing_solution} leaves a state $\widetilde\rho_S(t)$ with the same spectrum and coherence, whose matrix elements are
\begin{equation}
  [\widetilde\rho_S(t)]_{\bm s\bm s'}
  =\sqrt{w_{\bm s}w_{\bm s'}}\,
  \ee^{-\Gamma(t)\kappa_{\bm s\bm s'}},
  \label{eq:qrng_spatial_state}
\end{equation}
where the dimensionless dephasing coefficient for each pair of configurations is
\begin{equation}
  \kappa_{\bm s\bm s'}
  =\frac14\sum_{ij}(s_i-s'_i)
  \ee^{-|i-j|a/\xi}(s_j-s'_j).
  \label{eq:qrng_spatial_coefficient}
\end{equation}
Let $\mu_\alpha(\Gamma)$, $\alpha=1,\ldots,2^M$, denote the eigenvalues of this full matrix at fixed $M$, $q$, and $\xi/a$.  The finite-range reliability and resource hazard rate are then
\begin{equation}
  R_{{\rm RE},{\rm sp}}^{(q)}(t)
  =h_2(q)+\frac1M\sum_{\alpha=1}^{2^M}
  \mu_\alpha\log_2\mu_\alpha,
  \label{eq:qrng_spatial_reliability}
\end{equation}
\begin{equation}
  h_{{\rm RE},{\rm sp}}^{(q)}(t)
  =-\frac{\dot\Gamma(t)\sum_\alpha
  (\partial_\Gamma\mu_\alpha)\log_2\mu_\alpha}
  {\CRE[\rho_S(t)]}.
  \label{eq:qrng_spatial_hazard}
\end{equation}
Here $\mathrm{sp}$ denotes the spatially resolving bath, all eigenvalues are evaluated at $\Gamma=\Gamma(t)$, and $\sum_\alpha\partial_\Gamma\mu_\alpha=0$ was used in the second expression.  The rate is evaluated for $t>0$ while $\CRE[\rho_S(t)]>0$, with the same interpretation conditions as Eq.~\eqref{eq:qrng_hazard}.  These equations retain the full spatial kernel and reduce to the local formulas only when $C_{ij}=\delta_{ij}$.

For fixed finite $M$ and $0<\xi/a<\infty$, the exponential correlation matrix is positive definite.  Consequently $\kappa_{\bm s\bm s'}>0$ for distinct configurations.  Define the slowest configuration dephasing coefficient by
\begin{equation}
  \kappa_{\min}
  =\min_{\bm s\ne\bm s'}\kappa_{\bm s\bm s'}>0.
  \label{eq:qrng_spatial_minimum}
\end{equation}
As $\Gamma(t)\to\infty$, every off-diagonal element vanishes and $\rho_S(t)$ approaches the diagonal product distribution.  Since all $w_{\bm s}>0$ for $0<q<1$, the leading entropy deficit about this state is a positive quadratic form in the off-diagonal elements.  The terms with $\kappa_{\bm s\bm s'}=\kappa_{\min}$ therefore give
\begin{equation}
  R_{{\rm RE},{\rm sp}}^{(q)}(t)
  =A_q\,\ee^{-2\kappa_{\min}\Gamma(t)}[1+o(1)],
  \label{eq:qrng_spatial_asymptotic}
\end{equation}
where $A_q>0$ is a time-independent coefficient fixed by $M$, $q$, and $C_{ij}$, and $o(1)$ denotes a correction tending to zero.  Thus $h_{{\rm RE},{\rm sp}}^{(q)}(t)/\dot\Gamma(t)\to2\kappa_{\min}$.  For the Ohmic spectrum at finite nonzero temperature, the low-frequency behavior $\coth(\beta\hbar\omega/2)\sim2/(\beta\hbar\omega)$ and $\int_0^\infty(\sin x)/x\,\dd x=\pi/2$ give
\begin{equation}
  \dot\Gamma(t)\xrightarrow{t\to\infty}
  \frac{\pi\eta}{\beta\hbar}.
  \label{eq:qrng_ohmic_asymptotic_rate}
\end{equation}
It follows that
\begin{equation}
  h_{{\rm RE},{\rm sp}}^{(q)}(t)\xrightarrow{t\to\infty}
  \frac{2\pi\eta\kappa_{\min}}{\beta\hbar}.
  \label{eq:qrng_spatial_hazard_limit}
\end{equation}
In the strictly local limit $\kappa_{\min}=1$, giving $h_{{\rm RE},{\rm loc}}^{(q)}(\infty)=2\pi\eta/(\beta\hbar)$.  This nonzero limiting resource hazard rate describes the fractional loss of a resource that tends to zero.  The collective bath instead retains a nonzero resource and has $h_{\rm RE}^{(q)}(t)\to0$.  The finite-range asymptotics assume fixed finite $\xi/a$ and do not apply directly to the exactly collective matrix $C_{ij}=1$.

For the three-spin comparison below, minimizing Eq.~\eqref{eq:qrng_spatial_coefficient} over distinct configurations yields
\begin{equation}
  \kappa_{\min}
  =\min\!\left\{1,\,2[1-\ee^{-a/\xi}]\right\},
  \qquad M=3.
  \label{eq:qrng_three_spin_minimum}
\end{equation}
The two candidates correspond to changing one spin and to exchanging two opposite neighboring spins.  At $\xi/a=0.75$, the second coefficient is approximately $1.47281$, so $\kappa_{\min}=1$.  For $q=1/2$, expanding $S$ about $\mathbb{I}/8$, where $\mathbb{I}$ is the identity on the three-spin Hilbert space, and counting the 12 unordered pairs that differ at one spin gives
\begin{equation}
  R_{{\rm RE},{\rm sp}}^{(1/2)}(t)
  \sim\frac{\ee^{-2\Gamma(t)}}{2\ln2}.
  \label{eq:qrng_three_spin_asymptotic}
\end{equation}
This explains why the specified finite-range bath has the same limiting resource hazard rate as a local bath, although their finite-time reliabilities generally differ.

\textit{Numerical comparisons.} The numerical calculations test the analytical distinction above under a controlled comparison.  For each chosen $M$ and $q$, the two bath models use the same product preparation, diagonal spectrum $J_{ii}(\omega)=J(\omega)$, and isolated-spin dephasing kernel $\gamma_{ii}(t)=\Gamma(t)$.  Differences between the collective and spatially resolving results therefore arise from the off-diagonal spatial correlations, which determine both collective dephasing and bath-induced joint phases.

\begin{figure}[t]
  \centering
  \includegraphics[width=\columnwidth]{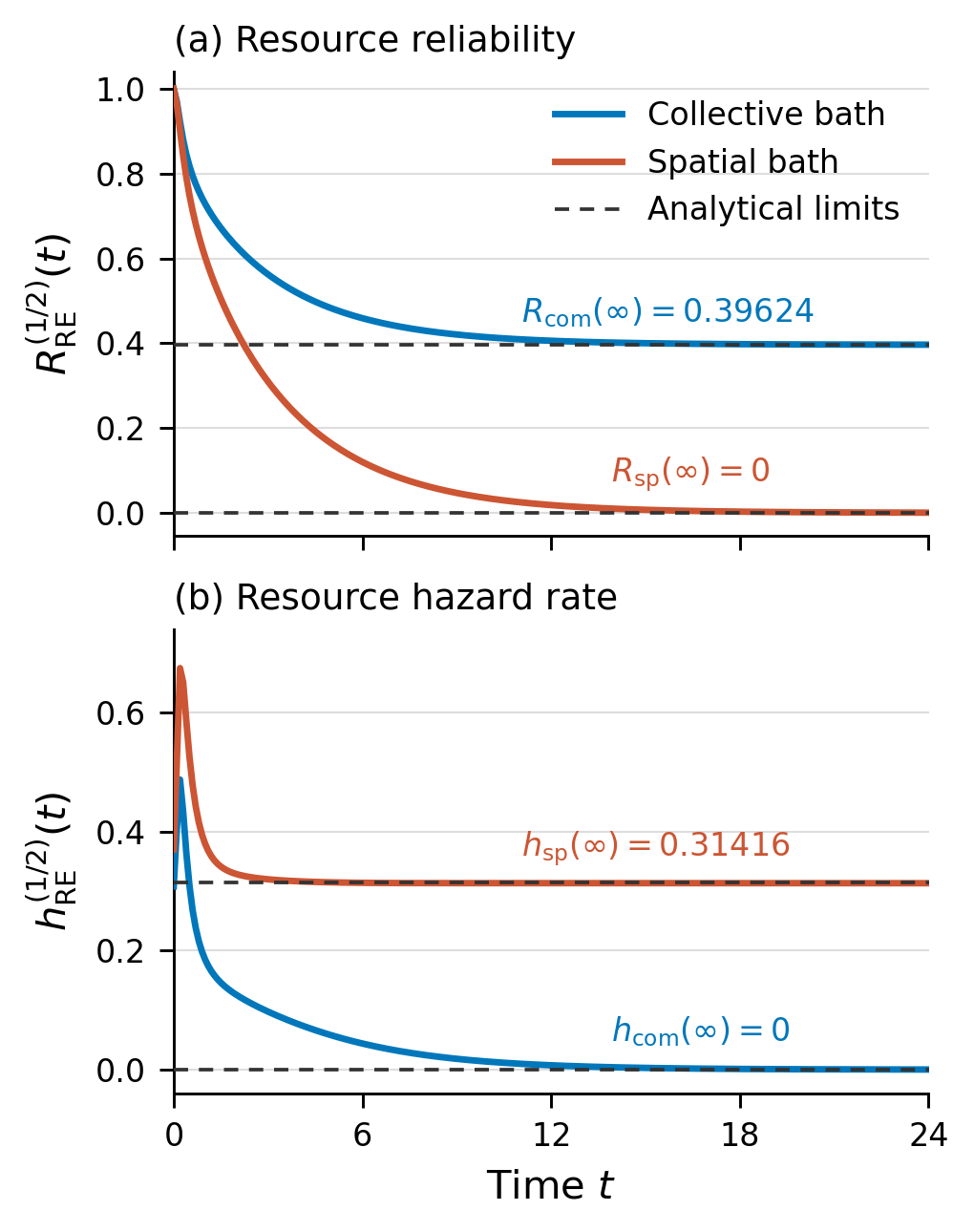}
  \caption{QRNG resource reliability with analytical long-time reference lines for $M=3$, $q=1/2$, $\eta=0.10$, $\omega_c=4$, $\beta=2$, and $\hbar=1$.  Both baths have the same isolated-spin kernel $\Gamma(t)$, and the spatial bath uses $C_{ij}=\exp[-|i-j|a/\xi]$ with $\xi/a=0.75$.  (a) Reliability curves and their limits $(\log_2 3)/4\simeq0.39624$ (collective) and $0$ (spatial).  (b) Numerically differentiated resource hazard rates and their limits $0$ (collective) and $2\pi\eta/\beta\simeq0.31416$ (spatial).  Solid curves are numerical results, and dashed lines are analytical predictions.  The annotations suppress the common $\mathrm{RE}$ and $(1/2)$ indices.}
  \label{fig:qrng_reliability}
\end{figure}

\begin{figure}[t]
  \centering
  \includegraphics[width=\columnwidth]{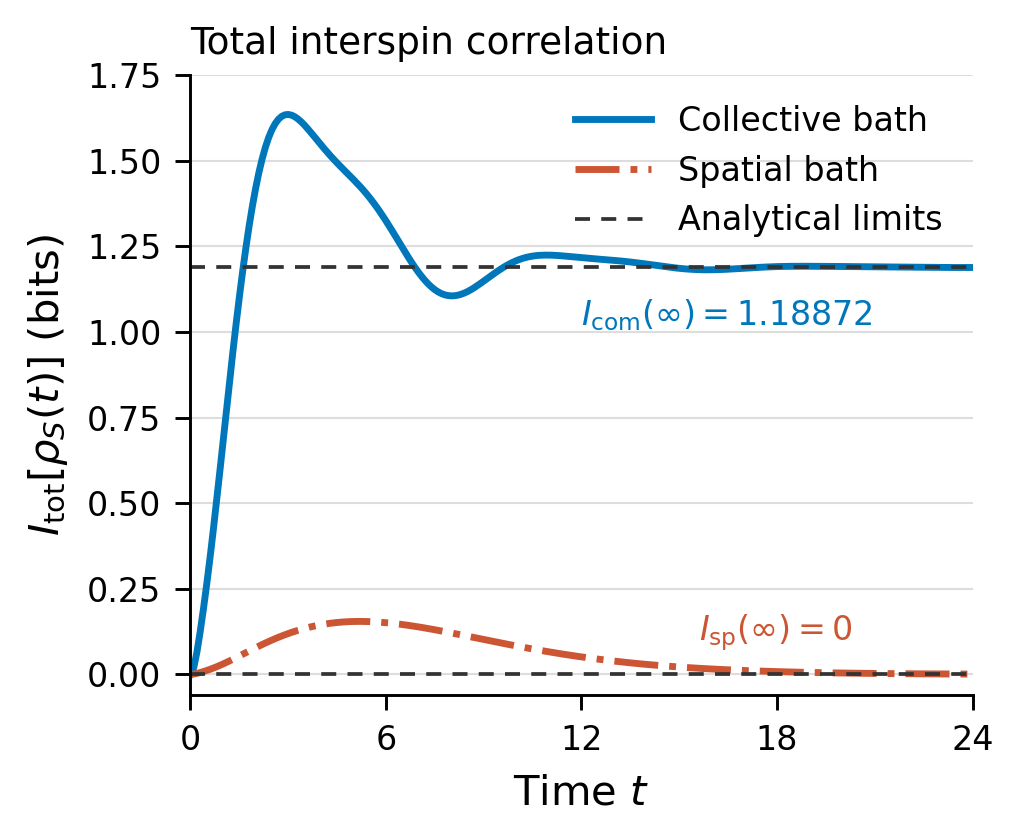}
  \caption{Total interspin correlation $I_{\rm tot}[\rho_S(t)]$ of the full reduced spin state, with all bath-induced phases retained.  Parameters are the same as in Fig.~\ref{fig:qrng_reliability}, with $M=3$ and $q=1/2$.  The solid blue and dash-dotted orange curves correspond to collective and finite-range spatial baths.  Dashed horizontal lines mark the analytical limits $I_{\rm com}(\infty)=3(\log_2 3)/4\simeq1.18872$ bits and $I_{\rm sp}(\infty)=0$, where $\mathrm{com}$ and $\mathrm{sp}$ label the two baths.  Equation~\eqref{eq:qrng_coherence_correlation_balance} relates this total correlation to the difference between total coherence and the sum of one-spin coherences.}
  \label{fig:qrng_bath_record}
\end{figure}

\begin{figure*}[t]
  \centering
  \includegraphics[width=\textwidth]{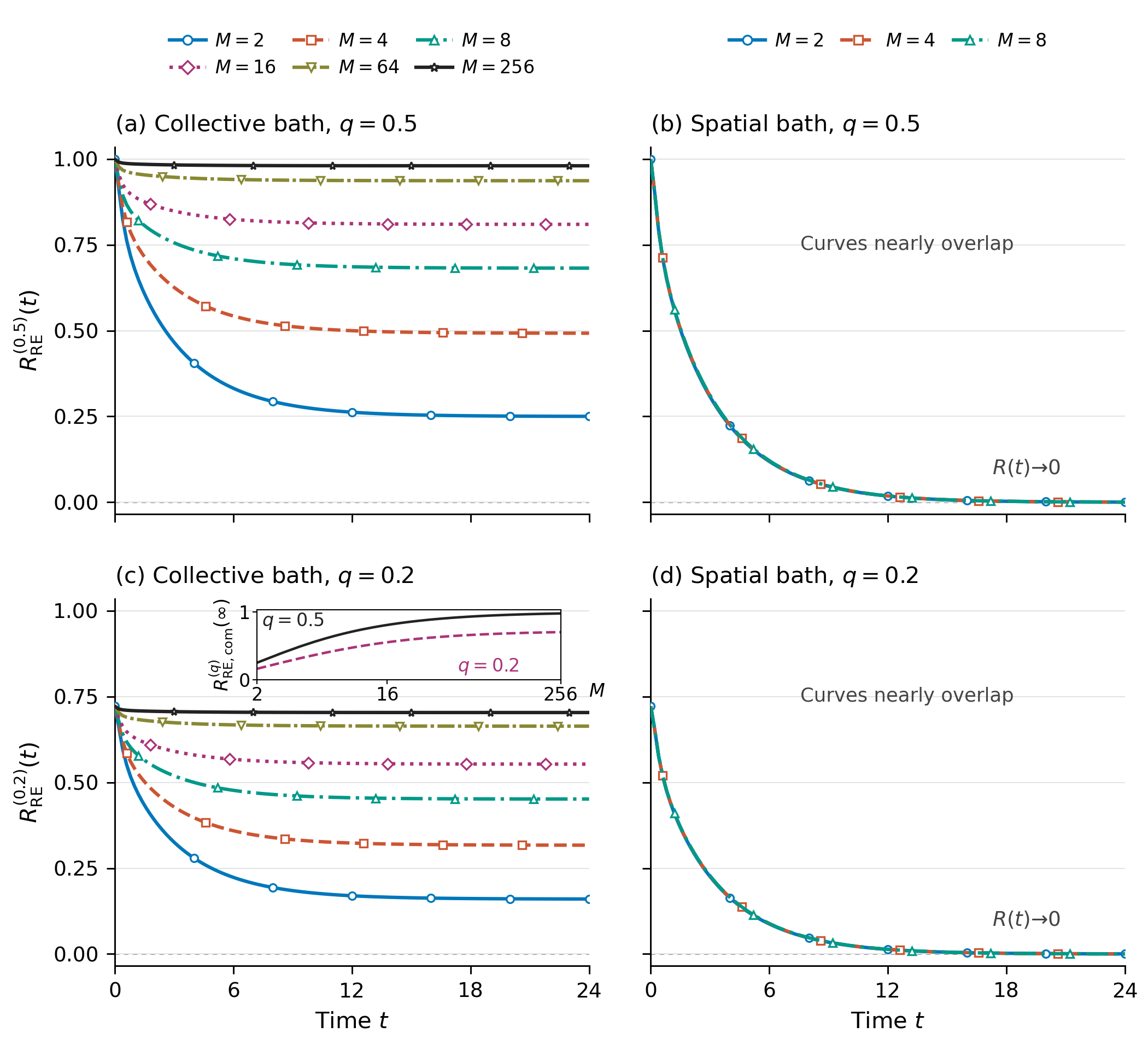}
  \caption{Register-size and preparation-bias dependence of QRNG resource reliability.  (a),(c) Collective bath with $M=2,4,8,16,64,256$.  (b),(d) Spatially resolving bath with $M=2,4,8$ and $\xi/a=0.75$.  The upper row uses $q=0.5$ and the lower row $q=0.2$.  Each column has its own size legend, with matching colors, line styles, and markers for the shared sizes.  Markers indicate selected numerical time points.  The common vertical scale displays $\CRE[\rho_S(t)]/M$, whose initial values are $1$ and $h_2(0.2)\simeq0.72193$, respectively.  Inset in (c): exact collective long-time reliability for $2\leq M\leq256$, evaluated from Eq.~\eqref{eq:qrng_compression_scaling} for both biases.  All bath parameters are the same as in Fig.~\ref{fig:qrng_reliability}.}
  \label{fig:qrng_compression}
\end{figure*}

We set the reduced Planck constant $\hbar=1$.  With the notation of Sec.~\ref{sec:spinboson}, $\eta$ is the dimensionless spin--bath coupling strength, $\omega_c$ is the Ohmic cutoff frequency, $\beta=(k_B T)^{-1}$ is the inverse bath temperature, $a$ is the lattice spacing, and $\xi/a$ is the spatial correlation length in lattice-spacing units.  For Figs.~\ref{fig:qrng_reliability}--\ref{fig:qrng_compression}, we use $\eta=0.10$, $\omega_c=4$, and $\beta=2$.  The spatially resolving calculation uses the full kernel $C_{ij}=\exp[-|i-j|a/\xi]$ with $\xi/a=0.75$.  The bath-record comparison and its correlation diagnostic use $M=3$ and $q=1/2$.  The size and bias comparison uses $q=0.5,0.2$, with collective registers up to $M=256$ and spatial registers of sizes $M=2,4,8$.

Figure~\ref{fig:qrng_reliability}(a) compares the resource reliabilities for $M=3$.  Both bath models begin at $\RRE^{(1/2)}(0)=1$.  The collective bath preserves coherence within each magnetization sector and approaches $R_{{\rm RE},{\rm com}}^{(1/2)}(\infty)=(\log_2 3)/4\simeq0.39624$, obtained from the exact binomial entropy in Eq.~\eqref{eq:qrng_compression_scaling}.  The finite-range spatial bath distinguishes those configurations and has $R_{{\rm RE},{\rm sp}}^{(1/2)}(\infty)=0$.  The horizontal dashed lines mark these analytical limits.

Figure~\ref{fig:qrng_reliability}(b) shows the corresponding resource hazard rates, evaluated as $-\dd\ln R/\dd t$ on the sampled time grid.  Equations~\eqref{eq:qrng_common_hazard_limit} and~\eqref{eq:qrng_spatial_hazard_limit} predict $h_{{\rm RE},{\rm com}}^{(1/2)}(\infty)=0$ and $h_{{\rm RE},{\rm sp}}^{(1/2)}(\infty)=2\pi\eta/\beta\simeq0.31416$ for these parameters and $\hbar=1$, also shown as dashed lines.  A vanishing reliability therefore need not have a vanishing resource hazard rate.  Together, the two panels show that the same isolated-spin attenuation $\ee^{-\Gamma(t)}$ can accompany either a protected many-spin resource plateau or continuing resource depletion.  Both trajectories satisfy $R(t)>0$ and $\dot R(t)\leq0$, so their resource hazard rates are nonnegative, consistently with Eq.~\eqref{eq:qrng_hazard}.

The total interspin correlation is
\begin{equation}
  I_{\rm tot}[\rho_S(t)]=\sum_{i=1}^M S[\rho_i(t)]-S[\rho_S(t)],
  \label{eq:qrng_total_correlation}
\end{equation}
where $\rho_i(t)$ is the one-spin marginal obtained by tracing out all other spins.  This quantity includes classical and quantum correlations.  It is neither an entanglement measure nor the mutual information between spins and bath.

For the product preparation in Eq.~\eqref{eq:biased_product_state}, pure dephasing preserves factorized readout-basis populations.  Hence $S[\Delta(\rho_S(t))]=M h_2(q)$ and $\CRE[\rho_i(t)]=h_2(q)-S[\rho_i(t)]$, with each one-spin coherence evaluated in its local readout basis.  Combining these expressions with Eqs.~\eqref{eq:qrng_reliability} and~\eqref{eq:qrng_total_correlation} gives
\begin{equation}
  \begin{aligned}
    M\RRE^{(q)}(t)&=\CRE[\rho_S(t)]\\
    &=\sum_{i=1}^M\CRE[\rho_i(t)]+I_{\rm tot}[\rho_S(t)].
  \end{aligned}
  \label{eq:qrng_coherence_correlation_balance}
\end{equation}
Thus the total correlation accounts for the coherence not captured by the sum of one-spin coherences.  Under strictly independent local evolution, $I_{\rm tot}=0$ for this preparation.  In the collective long-time limit, each one-spin coherence vanishes, but the surviving interspin correlation divided by $M$ equals the nonzero reliability plateau.  This decomposition relies on factorized readout populations and is not an identity for arbitrary states.

Figure~\ref{fig:qrng_bath_record} evaluates $I_{\rm tot}[\rho_S(t)]$ for the full reduced state, retaining the phase in Eq.~\eqref{eq:full_dephasing_phase} before taking the partial traces.  The collective-bath correlation first exceeds its long-time limit and then relaxes nonmonotonically to $3(\log_2 3)/4\simeq1.18872$ bits.  The finite-range spatial bath produces transient correlations that eventually vanish.  The bath-induced joint phases leave the reliability unchanged, but alter the one-spin entropies and hence the division between local coherence and total correlation in Eq.~\eqref{eq:qrng_coherence_correlation_balance}.  The nonmonotonic correlation curve therefore does not imply resource backflow in Fig.~\ref{fig:qrng_reliability}.  This diagnostic quantifies a contribution to the existing coherence resource, without introducing a separate reliability resource or directly measuring the environmental record.

Figure~\ref{fig:qrng_compression} extends the comparison to register size and preparation bias.  Panels (a) and (b) use $q=0.5$, while panels (c) and (d) use $q=0.2$.  The collective panels show $M=2,4,8,16,64,256$, whereas the spatial panels show $M=2,4,8$.  The shared sizes permit direct comparison between the baths, and the larger collective registers reveal the size-dependent protection.  The collective curves are evaluated from the $(M+1)$-dimensional compressed matrix $B(t)$, which permits the larger size range.  The spatial calculation instead diagonalizes the full $2^M$-dimensional reduced state with the specified $C_{ij}$.

In the collective bath, panels (a) and (c) show that increasing $M$ raises the retained reliability for both inputs.  At $M=256$, the reliability remains close to its preparation value throughout the plotted interval, not only in the long-time limit.  In the spatial bath, panels (b) and (d) show nearly overlapping curves for the sizes and correlation length considered.  This weak size dependence is not the exact $M$ independence of the strictly local result in Eq.~\eqref{eq:qrng_local_reliability}.  At fixed finite $M$ and $\xi$, the spatial reliability tends to zero as $\Gamma(t)\to\infty$, in contrast to the nonzero collective plateau.

The inset in panel (c) isolates the collective long-time size dependence.  It plots the exact limit $R_{{\rm RE},{\rm com}}^{(q)}(\infty)=h_2(q)-H_{\rm binomial}(M,q)/M$ from Eq.~\eqref{eq:qrng_compression_scaling}, rather than the value at the end of the finite-time grid.  For either fixed bias, the loss relative to the initial reliability decreases as $O(\log M/M)$.  Thus the long-time reliability approaches one for $q=0.5$ and $h_2(0.2)$ for $q=0.2$ as $M$ grows.  Collective noise protects an increasing fraction of the initially supplied resource, without raising reliability above its preparation value.  The contrast with the spatial bath shows that this protection depends on the collective coupling structure, not on register size alone.

The numerical conclusion also specifies the required trust assumption for model-assisted QRNG assessment.  A single-spin calibration determines $\Gamma(t)$, but it cannot determine the off-diagonal kernels $\gamma_{ij}(t)$ for $i\ne j$.  Inferring Eq.~\eqref{eq:common_qrng_reliability} from that calibration therefore additionally requires an independently validated collective-record model.  Otherwise, as the comparison shows, a bath model can fit the isolated-spin decay while giving a wrong multi-spin reliability.  Section~\ref{sec:assessment} makes this point with a finite-record stress test using $M=8$, $\eta=0.06$, $\omega_c=5$, and $\beta=3$.  There the common-record weight in Eq.~\eqref{eq:partial_record_kernel} is $c=1$ for the calibrated model and $c=0.65$ for the deliberately mismatched record.  That section also explains how a finite-block distillation benchmark reports an achievable yield together with its conversion error.

\section{Quantum interferometer}
\label{sec:interferometer}

\subsection{Device operation and resource reliability}

Unlike the QRNG in Sec.~\ref{sec:qrng}, the interferometer uses coherence to convert an encoded phase into a change in output probabilities.  We consider a spin implementation of the preparation, phase-accumulation, and analysis sequence used in Ramsey interferometry~\cite{giovannetti2006quantum,pezze2018rmp}.  The states $\ket{0}$ and $\ket{1}$ label the two paths of one spin.  Starting from $\ket{0}^{\otimes M}$, the control Hamiltonians are
\begin{align}
  H_S^{(1)}&=\frac{\hbar\Omega_1}{2}\sum_i\sigma_i^x,
  &&0<u<T_1,\\
  H_S^{(2)}&=\frac{\hbar\omega_z}{2}\sum_i\sigma_i^z,
  &&T_1<u<T_1+T_2,\\
  H_S^{(3)}&=-\frac{\hbar\Omega_3}{2}\sum_i\sigma_i^x,
  &&T_1+T_2<u<T_1+T_2+T_3.
  \label{eq:interferometer_controls}
\end{align}
Here $u$ is the elapsed time of the full sequence, $T_j$ is the duration of stage $j=1,2,3$, and $\Omega_1,\Omega_3$ are positive pulse angular frequencies.  Ideal preparation and inverse-analysis pulses satisfy $\Omega_1T_1=\Omega_3T_3=\pi/2$.  The first pulse creates path coherence, the second stage encodes a relative phase, and the third pulse converts that phase into a $\sigma_z$ population signal.  Known local phase rotations relate this convention to $\ket{+}^{\otimes M}$ preparation and Hadamard readout below.

The bath Hamiltonian and longitudinal coupling remain those of Sec.~\ref{sec:spinboson}.  Only the second-stage spin Hamiltonian commutes with the coupling.  In the analytical phase-stage benchmark, $t$ denotes time measured from the entrance to that stage, with $t=T_2$ at readout, and $\rho_S(0)\otimes\rho_B$ is assumed factorized there.  Ideal short pulses or a separately prepared, decorrelated input justify this reference setting.  Finite pulses can generate spin--bath correlations, so the subsequent dynamics cannot in general be obtained by restarting Eq.~\eqref{eq:exact_dephasing_solution} with the reduced spin state alone.

We consider a lossless, calibrated two-port task.  Every use produces one recorded outcome, without postselecting successful detections.  For one spin with ideal analysis, the two output probabilities are
\begin{equation}
  \begin{aligned}
    P_0(\phi,t)&=\frac12[1+V(t)\cos(\phi+\phi_0)],\\
    P_1(\phi,t)&=1-P_0(\phi,t),
  \end{aligned}
  \label{eq:ramsey_fringe}
\end{equation}
where $\phi$ is the encoded phase, including $\omega_z t$ and any applied phase shift, $\phi_0$ is a calibrated offset, and $V(t)\in[0,1]$ is the visibility.  A phase scan is performed at fixed $t$ and fixed noise parameters.  The factorized pure-dephasing benchmark gives $V(t)=V(0)\ee^{-\Gamma(t)}$.  The bath-response phase in Eq.~\eqref{eq:full_dephasing_phase} cancels for a single spin, but not generally for multiple spins.

Distinguishable path records constrain the attainable fringe visibility~\cite{englert1996visibility}. In the path basis, the $l_1$ coherence is $C_{l_1}(\rho)=\sum_{a\ne b}|\rho_{ab}|$, where $a,b$ label basis states.  For a balanced one-spin state immediately before ideal analysis,
\begin{equation}
  \rho=
  \begin{pmatrix}
    1/2 & c\\
    c^* & 1/2
  \end{pmatrix},
  \qquad C_{l_1}(\rho)=2|c|=V,
  \label{eq:visibility_l1}
\end{equation}
where $c$ is its complex off-diagonal matrix element.  The equality with visibility follows from balanced two-path readout~\cite{bera2015}.  As in Sec.~\ref{sec:qrng_definition}, we use incoherent operations in the fixed reference basis as the free-operation class.  Preparation and analysis are calibrated resource-creating or resource-converting controls, not free noise operations.  The logarithmic coherence~\cite{rana2016logarithmic}
\begin{equation}
  \Clog(\rho)=\log_2[1+C_{l_1}(\rho)]
  \label{eq:log_l1_def}
\end{equation}
is a coherence monotone under these operations and is additive on independent states $\rho$ and $\sigma$:
\begin{equation}
  \Clog(\rho\otimes\sigma)=\Clog(\rho)+\Clog(\sigma).
  \label{eq:log_l1_additivity}
\end{equation}
Its use here is based on this composition rule and the calibrated visibility relation, not on the coherent-qubit distillation rate used for the QRNG.

For the one-spin operational sector, $\Clog^{\max}=1$.  Equation~\eqref{eq:resource_reliability} therefore gives
\begin{equation}
  \Rfr(t)=\Clog[\rho_S(t)]=\log_2[1+V(t)],
  \label{eq:interferometer_reliability}
\end{equation}
where $\rho_S(t)$ is evaluated before analysis and $\mathrm{fr}$ denotes the fringe resource.  The normalization refers to the ideal unit, so imperfect preparation may give $\Rfr(0)<1$.  Detector loss, unequal efficiencies, pulse errors, and phase drift require a calibrated measurement model before an observed contrast can be identified with this state resource.  They cannot be absorbed into $V$ while retaining Eq.~\eqref{eq:visibility_l1} without further justification.  Section~\ref{sec:assessment} describes the ideal-model count estimator and its calibration requirements.

Applying Eq.~\eqref{eq:resource_hazard} on positive, nonincreasing reliability intervals yields
\begin{equation}
  \begin{aligned}
    h_{\rm fr}(t)
    &=-\frac{\dot V(t)}{[1+V(t)]\ln[1+V(t)]}\\
    &=\frac{\dot\Gamma(t)V(t)}
    {[1+V(t)]\ln[1+V(t)]},
  \end{aligned}
  \label{eq:interferometer_hazard}
\end{equation}
where the second equality assumes $V(t)=V(0)\ee^{-\Gamma(t)}$.  This is a resource hazard rate, not automatically a classical first-failure hazard.  If a revival makes $\dot V>0$, the same logarithmic derivative becomes negative.

Which output port is correct depends on the chosen phase operating point, so a given port does not always represent failure.  Resource reliability quantifies the coherence retained before readout, whereas the correct-port probability describes the chance of obtaining the expected outcome in one use.  These quantities are related but distinct.

After calibrating $\phi_0=0$, choose port $0$ at the bright fringe $\phi=0$ and port $1$ at the dark fringe $\phi=\pi$.  The correct-port probability is
\begin{equation}
  p_{\rm c}(t)=\frac{1+V(t)}{2}=2^{\Rfr(t)-1}.
  \label{eq:correct_port_probability}
\end{equation}
Thus zero coherence gives random guessing, $p_{\rm c}=1/2$, rather than a zero probability of a correct output.  For $N_{\rm use}$ independent, identically prepared repetitions at the same phase-stage duration $t$,
\begin{align}
  P_{\rm all}^{(N_{\rm use})}(t)
  &=[p_{\rm c}(t)]^{N_{\rm use}}
  =\left(\frac{1+V(t)}{2}\right)^{N_{\rm use}}
  \nonumber\\
  &=2^{-N_{\rm use}[1-\Rfr(t)]}.
  \label{eq:all_correct_probability}
\end{align}
Here $N_{\rm use}$ counts complete device uses.  At the bright fringe this is the probability of the string $0^{N_{\rm use}}$, and at the dark fringe it is that of $1^{N_{\rm use}}$.  The product rule requires independent environments or effective bath reset between uses and no shared preparation or calibration fluctuations.  Memory within one use does not invalidate the rule, but inter-use correlations do.

The last equality follows directly from $\Rfr(t)=\log_2[1+V(t)]$.  Under the stated independent-use conditions, resource reliability therefore determines the all-correct event probability, but is not itself that probability.

\subsection{Analytical expressions and numerical comparisons}

\textit{Coherent-spin input.} A coherent spin state (CSS) with identical equatorial spins is locally phase-equivalent to $\ket{+}^{\otimes M}$, the $q=0.5$ product input of Sec.~\ref{sec:qrng_model_results}.  Its ideal logarithmic coherence is $M$, so the corresponding normalized resource benchmark is
\begin{equation}
  R_{{\rm fr},{\rm CSS}}^{(M)}(t)
  =\frac{\Clog[\rho_{\rm CSS}(t)]}{M}.
  \label{eq:css_fringe_reliability}
\end{equation}
For the collective bath, group configurations by the number $n$ of spins in $\ket{1}$.  Each sector has $\binom{M}{n}$ configurations, and each matrix element has initial magnitude $2^{-M}$.  Summing their dephased magnitudes gives
\begin{equation}
  \begin{aligned}
    R_{{\rm fr},{\rm CSS},{\rm com}}^{(M)}(t)
    &=\frac1M\log_2\!\Bigg[
    2^{-M}\sum_{n,n'=0}^M\\
    &\quad\times\binom{M}{n}\binom{M}{n'}
    \ee^{-\Gamma(t)(n-n')^2}\Bigg].
  \end{aligned}
  \label{eq:css_collective_reliability}
\end{equation}
The bath-induced diagonal unitary does not change these magnitudes.  Unlike the entropy in Sec.~\ref{sec:qrng_model_results}, $l_1$ coherence is not determined by the compressed-state spectrum alone.  The sector multiplicities must be retained.

When $\Gamma(t)\to\infty$, only $n=n'$ contributes.  Using $\sum_n\binom{M}{n}^2=\binom{2M}{M}$ and then Stirling's approximation gives
\begin{equation}
  \begin{aligned}
    R_{{\rm fr},{\rm CSS},{\rm com}}^{(M)}(\infty)
    &=\frac{\log_2\binom{2M}{M}}{M}-1\\
    &=1-\frac{\log_2(\pi M)}{2M}+O(M^{-2}).
  \end{aligned}
  \label{eq:css_collective_limit}
\end{equation}
The expansion assumes large $M$.  In the strictly local bath, tensor-product evolution and Eq.~\eqref{eq:log_l1_additivity} instead give
\begin{equation}
  R_{{\rm fr},{\rm CSS},{\rm loc}}^{(M)}(t)
  =\log_2[1+\ee^{-\Gamma(t)}].
  \label{eq:css_local_reliability}
\end{equation}
This expression is independent of $M$ and tends to zero.  A finite-range spatial kernel requires the full configuration sum and also gives zero at complete dephasing for fixed finite $M$ and $\xi$.

The nonzero collective plateau is a statement about stored basis coherence, not a plateau of sensitivity to the common phase.  At complete collective dephasing, the remaining coherences connect states with the same eigenvalue of $J_z=\frac12\sum_i\sigma_i^z$.  The state then commutes with $J_z$ and carries no information about the phase generated by $J_z$.  Consequently, Eq.~\eqref{eq:css_fringe_reliability} is a resource benchmark for the CSS mode, not a universal measure of its phase-estimation performance.  That performance must be checked using the specified encoding and readout.

\textit{GHZ input.} A Greenberger--Horne--Zeilinger (GHZ) state,
\begin{equation}
  \ket{\rm GHZ}=
  \frac{\ket{0}^{\otimes M}+\ket{1}^{\otimes M}}{\sqrt2},
  \label{eq:ghz_input}
\end{equation}
requires an entangling preparation circuit, not the independent first-stage rotations in Eq.~\eqref{eq:interferometer_controls}.  Its operational sector consists of these two branches and has $\Clog^{\max}=1$.  For a real correlation matrix $C$, Eq.~\eqref{eq:general_dephasing_exponent} gives
\begin{equation}
  \mathcal V_{\rm GHZ}(t)
  =\exp\!\left[-\Gamma(t)\sum_{i,j=1}^M C_{ij}\right].
  \label{eq:ghz_general_visibility}
\end{equation}
The bath-response phase cancels because the two opposite spin configurations have equal $\bm s^{\mathsf T}C\bm s$.  For the exponential kernel,
\begin{equation}
  \sum_{i,j=1}^M C_{ij}
  =M+2\sum_{r=1}^{M-1}(M-r)\ee^{-ra/\xi},
  \label{eq:ghz_spatial_factor}
\end{equation}
where $r$ is the separation in lattice sites.  In particular,
\begin{equation}
  \mathcal V_{\rm GHZ}(t)=
  \begin{cases}
    \ee^{-M\Gamma(t)}, & \text{local bath},\\
    \ee^{-M^2\Gamma(t)}, & \text{collective bath}.
  \end{cases}
  \label{eq:ghz_visibility_scaling}
\end{equation}
The branch-resource reliability is therefore
\begin{equation}
  R_{{\rm fr},{\rm GHZ}}(t)=\log_2[1+\mathcal V_{\rm GHZ}(t)].
  \label{eq:ghz_fringe_reliability}
\end{equation}
The CSS and GHZ normalizations refer to different operational sectors and do not compare equal absolute amounts of coherence.

Ideal parity analysis gives
\begin{equation}
  P_{\rm even}^{\rm GHZ}(\phi,t)
  =\frac12[1+\mathcal V_{\rm GHZ}(t)\cos(M\phi)],
  \label{eq:ghz_parity_fringe}
\end{equation}
where an even outcome contains an even number of $\ket{1}$ results.  Thus the larger phase response $M\phi$ is accompanied by enhanced longitudinal-dephasing sensitivity~\cite{huelga1997improvement}.  At calibrated parity extrema, Eq.~\eqref{eq:correct_port_probability} applies with the two logical outputs even and odd.  A repeated-use product rule still requires independent complete GHZ preparations and bath resets.

\begin{figure}[!tb]
  \centering
  \includegraphics[width=\columnwidth]{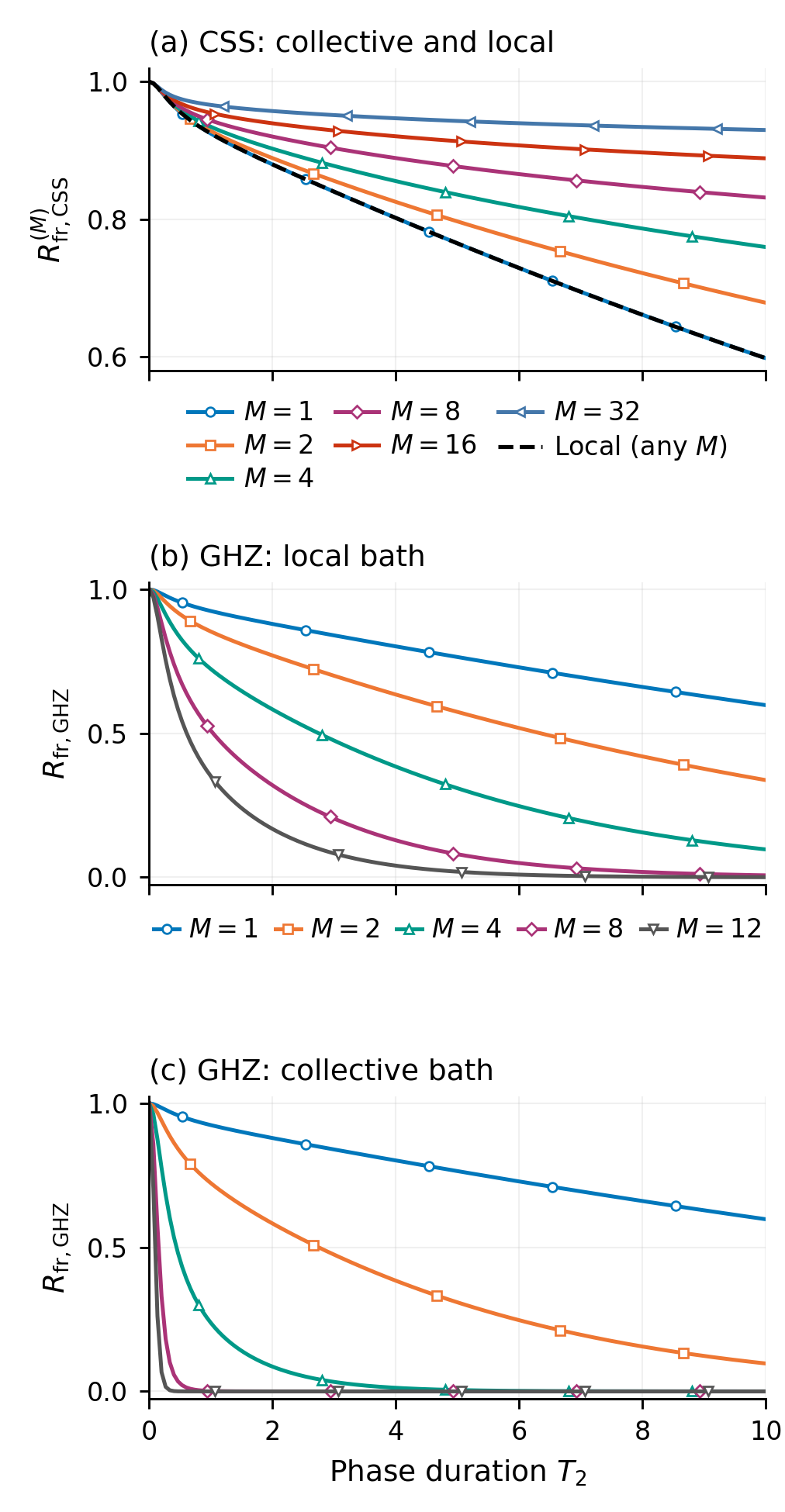}
  \caption{Normalized logarithmic-coherence resources during phase accumulation, with $\eta=0.06$, $\omega_c=5$, $\beta=3$, and $\hbar=1$.  (a) CSS reliability $\Clog/M$ in collective (colored solid curves) and local (black dashed curve) baths.  (b),(c) GHZ branch reliability $\Clog$ in local and collective baths, respectively.  Colors and markers identify $M$, with a shared legend for (b) and (c).  The CSS collective plateau concerns basis coherence and does not imply nonzero long-time sensitivity to a common phase.  The two input families use different sector normalizations.}
  \label{fig:interferometer_css_ghz}
\end{figure}

The resource-reliability curves for both input families are compared in Fig.~\ref{fig:interferometer_css_ghz}.

\textit{Readout probabilities.} To calculate the readout fringes, let $\bm b,\bm b',\bm x\in\{0,1\}^M$ denote binary strings.  Their associated spin eigenvalues are $s_i(\bm b)=(-1)^{b_i}$, and $g_{\bm b}=M/2-w(\bm b)$ is the eigenvalue of $J_z$, with $w(\bm b)=\sum_i b_i$.  Define
\begin{equation}
  \begin{aligned}
    D_{\bm b\bm b'}(t)
    &=\ee^{-\Gamma_{\bm s(\bm b)\bm s(\bm b')}(t)}\\
    &\quad\times\ee^{\,\ii[\vartheta_{\bm b}(t)-\vartheta_{\bm b'}(t)]},\\
    \vartheta_{\bm b}(t)
    &=\frac{\Theta(t)}4\,\bm s(\bm b)^{\mathsf T}C\bm s(\bm b).
  \end{aligned}
  \label{eq:interferometer_complex_kernel}
\end{equation}

\begin{figure}[!tb]
  \centering
  \includegraphics[width=\columnwidth]{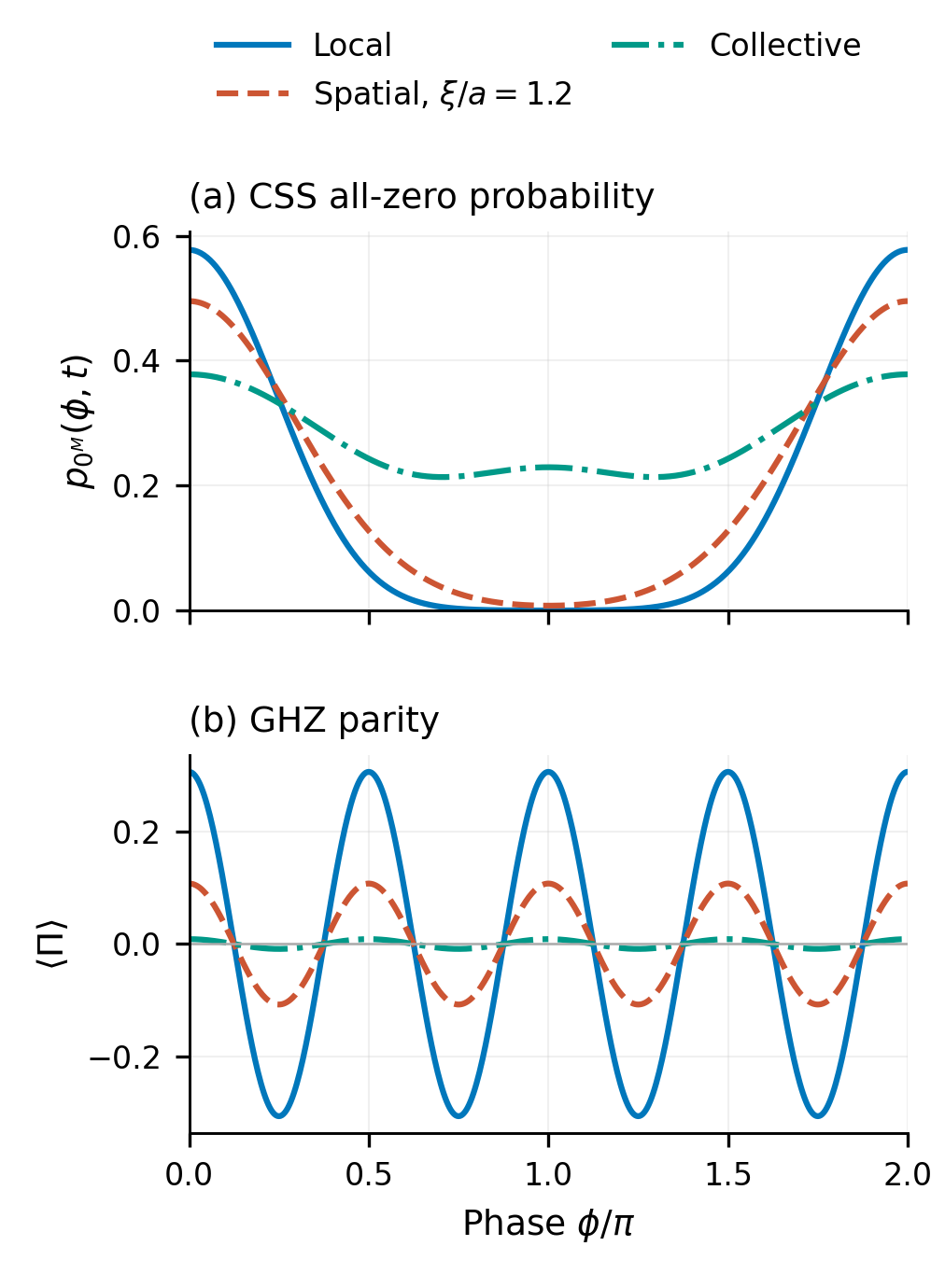}
  \caption{Full-phase interferometer readout for $M=4$, $T_2=4$, $\eta=0.06$, $\omega_c=5$, $\beta=3$, and $\hbar=1$.  Ideal preparation and analysis are assumed.  (a) CSS all-zero probability and (b) GHZ parity mean for local, spatial ($\xi/a=1.2$), and collective baths at the same $\Gamma(t)$.  All bath-induced phases are retained.}
  \label{fig:interferometer_readout}
\end{figure}
Here $\vartheta_{\bm b}$ is the bath-induced phase of configuration $\bm b$.  This factor retains the bath response from Eq.~\eqref{eq:full_dephasing_phase}.  Free precession is included once, in the scanned phase $\phi$.  For the Hadamard gate $H_{\rm Had}=(\sigma_x+\sigma_z)/\sqrt2$, ideal $H_{\rm Had}^{\otimes M}$ analysis gives
\begin{align}
  p_{\bm x}(\phi,t)
  &=2^{-M}\sum_{\bm b,\bm b'}
  (-1)^{\bm x\cdot(\bm b\oplus\bm b')}
  \rho_{\bm b\bm b'}(0)
  \nonumber\\
  &\quad\times D_{\bm b\bm b'}(t)
  \ee^{-\ii\phi(g_{\bm b}-g_{\bm b'})}.
  \label{eq:full_bitstring_probability}
\end{align}
Here $\oplus$ is bitwise addition modulo two, and the dot product in the sign is evaluated modulo two.  The matrix elements $\rho_{\bm b\bm b'}(0)$ refer to the spin state at the phase-stage entrance.  Hadamard analysis and the corresponding inverse equatorial rotation give the same readout probabilities up to calibrated phase conventions.

\textit{Numerical comparisons.} We compare phase-stage resource reliability and readout using the continuum Ohmic kernels with $\hbar=1$, $\eta=0.06$, $\omega_c=5$, and $\beta=3$.  Within every bath comparison, $C_{ii}=1$ fixes the same isolated-spin attenuation.  The resource curves use the collective and strictly local limits.  The readout figure additionally uses $C_{ij}=\exp[-|i-j|a/\xi]$ with $\xi/a=1.2$.  These device-specific parameters differ from those in Sec.~\ref{sec:qrng_model_results} but use the same model definitions.

Figure~\ref{fig:interferometer_css_ghz} evaluates Eqs.~\eqref{eq:css_collective_reliability}, \eqref{eq:css_local_reliability}, and~\eqref{eq:ghz_fringe_reliability}.  In panel (a), the CSS collective curves retain more normalized coherence as $M$ increases, consistently with Eq.~\eqref{eq:css_collective_limit}.  Panels (b) and (c) show that GHZ reliability decreases more rapidly with $M$ in local and collective baths, respectively, with the collective exponent growing as $M^2$.  These resource curves are unaffected by the deterministic diagonal phases, but their readout implications depend on the encoding.

Figure~\ref{fig:interferometer_readout} evaluates the full complex kernel in Eq.~\eqref{eq:interferometer_complex_kernel}, without compensating its bath-induced phase.  All panels use $M=4$ and $t=T_2=4$.  Panels (a) and (b) compare the CSS all-zero probability and GHZ parity mean in local, finite-range, and collective baths.  The parity mean is $\langle\Pi\rangle=2P_{\rm even}-1$, where $\Pi=(-1)^{w(\bm x)}$ is the measured parity.  At the same isolated-spin attenuation, bath correlations change the CSS fringe shape and the GHZ parity amplitude.

\section{Finite-record assessment of resource reliability}
\label{sec:assessment}

Resource-reliability assessment estimates the capacity remaining after a specified storage or phase-accumulation time from repeated preparations and measurements.  For the two devices considered here, we first construct a visibility estimator from interferometer counts, then use the same counting protocol on a reference spin to infer QRNG reliability under a calibrated correlation model.  Repeated simulated records quantify the resulting sampling fluctuations and the effect of model mismatch.

For the lossless, calibrated one-spin interferometer of Sec.~\ref{sec:interferometer}, take ideal analysis and a known phase offset $\phi_0=0$.  At fixed phase-accumulation time $t$, the port-$0$ probabilities at the two phase settings are
\begin{align}
  p_+(t)&=P_0(0,t)=\frac{1+V(t)}{2},\\
  p_-(t)&=P_0(\pi,t)=\frac{1-V(t)}{2}.
\end{align}
With $N_s$ independent measurements at each setting, the port-$0$ counts $n_+$ and $n_-$ are independent binomial variables with parameters $(N_s,p_+)$ and $(N_s,p_-)$.  The full record contains $2N_s$ measurements.  Port $0$ is correct at the bright fringe and port $1$ at the dark fringe, giving a combined correct count $K=n_++N_s-n_-$.  Up to terms independent of $V$, the log likelihood is
\begin{equation}
  \begin{aligned}
    \ell(V)={}&K\ln\frac{1+V}{2}\\
             &+(2N_s-K)\ln\frac{1-V}{2},
    \qquad 0\leq V\leq1.
  \end{aligned}
  \label{eq:visibility_likelihood}
\end{equation}
Maximization over this physical interval gives
\begin{equation}
  \begin{aligned}
    \widetilde V&=\frac{n_+-n_-}{N_s},\\
    \widehat V&=\min\{1,\max\{0,\widetilde V\}\},\\
    \widehat R_{\rm fr}&=\log_2(1+\widehat V).
  \end{aligned}
  \label{eq:two_phase_estimator}
\end{equation}
The direct count-difference estimate $\widetilde V$ lies in $[-1,1]$.  The physical constraint therefore sets negative estimates to zero and leaves nonnegative estimates unchanged.  Binomial independence gives the exact moments
\begin{equation}
  \mathbb E[\widetilde V]=V,
  \qquad
  {\rm Var}(\widetilde V)=\frac{1-V^2}{2N_s}.
  \label{eq:visibility_estimator_variance}
\end{equation}
This variance also approximates that of $\widehat V$ when negative estimates have negligible probability.  Near $V=0$, setting them to zero introduces bias.  The nonlinear conversion from visibility to reliability introduces an additional finite-sample bias.  Both effects are retained in the simulations.

The same record yields an estimate of the probability that all $N_{\rm use}$ complete uses give the correct output:
\begin{equation}
  \widehat P_{\rm all}^{(N_{\rm use})}(t)
  =\left(\frac{1+\widehat V(t)}{2}\right)^{N_{\rm use}}.
  \label{eq:all_correct_estimator}
\end{equation}
Here $N_s$ is the assessment sample size and $N_{\rm use}$ is the number of complete uses specified by the task.  This inference retains the independent-use assumption of Eq.~\eqref{eq:all_correct_probability}.  Data at different $t$ come from fresh preparations.  Detector loss, unequal efficiencies, and phase drift require the corresponding calibration parameters to be included in the likelihood.

QRNG assessment requires the relative-entropy coherence in Eq.~\eqref{eq:cre_def}.  For a general many-spin state, this quantity can be calculated from a tomographically reconstructed state~\cite{nielsen2010quantum}.  Randomized measurements and classical shadows provide estimators for selected state properties~\cite{brydges2019probing,huang2020predicting,elben2023randomized}.  In particular, Brydges et al. estimate the second-order R\'enyi entropy, which is distinct from the von Neumann entropy entering $\CRE$.  The above works do not by themselves establish an efficient, general estimator of the von Neumann entropy required here.  Applying these methods to resource-reliability assessment requires a suitable entropy-estimation protocol and an analysis of its sampling cost and error.  A direct finite-block distillation test instead reports $N$, $M$, the target length $m_N$, the conversion error $\epsilon_N$, and the operations used.  The ratio $m_N/(NM)$ is the achievable target-slot probability of that protocol, whose optimal vanishing-error limit gives $\RRE(t)$ as in Eq.~\eqref{eq:qrng_yield_fraction}.

For the symmetric product input with known $q$ and $M$, an independently calibrated collective coupling structure permits inference from a reference spin.  A balanced reference spin with unit initial visibility has $V(t)=\ee^{-\Gamma(t)}$.  Its two-phase counts give $\widehat V(t)$ through Eq.~\eqref{eq:two_phase_estimator}, and hence
\begin{equation}
  \begin{aligned}
    \widehat\Gamma(t)&=-\ln\widehat V(t),\\
    \widehat R_{{\rm RE},{\rm com}}^{(q)}(t)
      &=h_2(q)-\frac{S[B(\widehat\Gamma(t))]}{M}.
  \end{aligned}
  \label{eq:model_assisted_qrng_estimator}
\end{equation}
The matrix $B(\widehat\Gamma)$ is obtained from Eq.~\eqref{eq:compressed_B} using the estimated kernel and known sector weights $p_n(q)$.  For $\widehat V=0$, we use the limit $\widehat\Gamma\to\infty$, in which $B$ is diagonal with entries $p_n(q)$.  The reference spin determines only the single-spin attenuation.  The collective coupling assumption requires independent correlation measurements, for which multiqubit noise spectroscopy can provide the relevant cross-correlation spectra~\cite{pazsilva2017multiqubit}.

To test the effect of an incorrect correlation model, we use the diagnostic family
\begin{equation}
  C_{ij}^{(c)}=c+(1-c)\delta_{ij},
  \qquad 0\leq c\leq1,
  \label{eq:assessment_correlation_matrix}
\end{equation}
where $c$ is the weight of the collective contribution.  This positive-semidefinite matrix combines collective and independent local noise while keeping $C_{ii}^{(c)}=1$.  Substituting it into Eq.~\eqref{eq:general_dephasing_exponent} and removing diagonal unitary phases, which leave $\CRE$ unchanged, gives the real attenuation factor
\begin{equation}
  \begin{aligned}
    \mathcal D^{(c)}_{\bm s\bm s'}(t)
    &=\exp\!\bigl[-\Gamma(t)A^{(c)}_{\bm s\bm s'}\bigr],\\
    A^{(c)}_{\bm s\bm s'}
    &=c(n_{\bm s}-n_{\bm s'})^2
      +(1-c)d_H(\bm s,\bm s').
  \end{aligned}
  \label{eq:partial_record_kernel}
\end{equation}
Here $n_{\bm s}=(M-\sum_i s_i)/2$ counts spins in $\ket1$, and $d_H(\bm s,\bm s')$ counts sites where the two configurations differ.  The limits $c=1$ and $c=0$ correspond to collective and local environments.  This diagnostic family is defined separately from the exponential spatial kernel in Sec.~\ref{sec:spinboson}.

\begin{figure*}[!t]
  \centering
  \includegraphics[width=\textwidth]{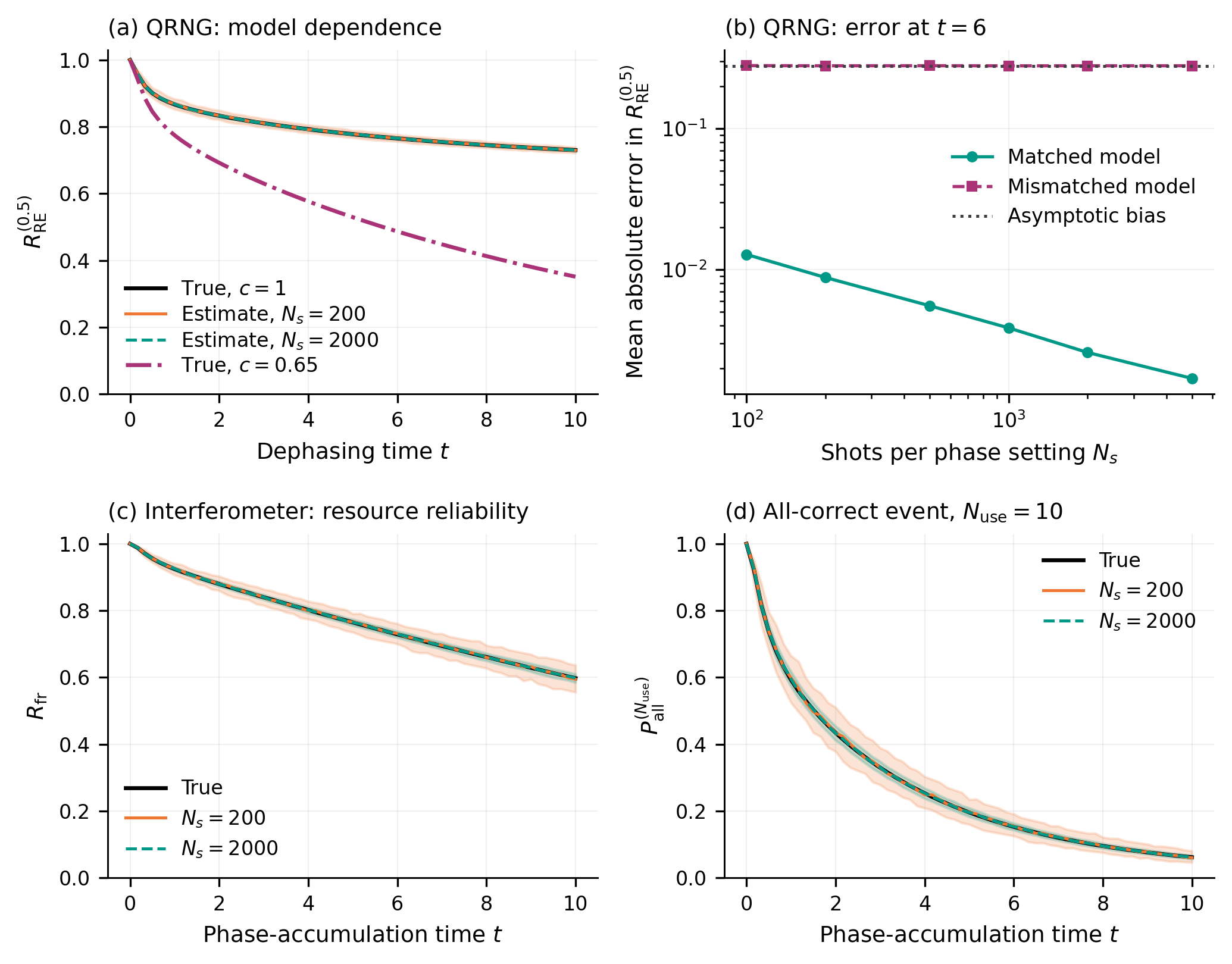}
  \caption{Finite-record resource-reliability assessment.  (a) QRNG inference assuming collective noise, compared with true collective ($c=1$) and partially correlated ($c=0.65$) resources at $M=8$, $q=0.5$, and identical isolated-spin attenuation.  (b) Mean absolute estimation error at $t=6$.  The dotted line is the asymptotic model-mismatch bias.  (c) One-spin interferometer reliability $R_{\rm fr}$.  (d) The inferred all-correct probability for $N_{\rm use}=10$ independent complete uses.  Parameters are $\eta=0.06$, $\omega_c=5$, $\beta=3$, and $\hbar=1$.  Colored curves in (a),(c),(d) are medians over 600 simulated records with $N_s$ shots per phase setting.  Shaded regions are pointwise 16th--84th percentile sampling bands, not experimental confidence intervals.}
  \label{fig:measurement_assessment}
\end{figure*}

All members of this family give the same isolated-spin attenuation $\ee^{-\Gamma(t)}$, while the many-spin reliability depends on $c$.  In the mismatch test, the true environment has $c=0.65$ and the inference assumes $c=1$.  Because the single-spin record is insensitive to $c$, additional probe measurements reduce sampling fluctuations but cannot remove the bias caused by the incorrect correlation model.  Figure~\ref{fig:measurement_assessment} compares the finite-record results for both devices.

The numerical assessment uses $\eta=0.06$, $\omega_c=5$, $\beta=3$, and $\hbar=1$.  Both the one-spin interferometer and the reference spin have $V(0)=1$, and the QRNG input has $M=8$ and $q=0.5$.  We select 61 times in $0\leq t\leq10$ and simulate 600 independent records for each sample size $N_s=200$ and $2000$ per phase setting.  The curves show estimator medians and pointwise 16th--84th percentile sampling bands.  The same simulated counts are passed through the two inference formulas to compare their statistical responses.

Panels (c) and (d) show the interferometer resource reliability and the all-correct probability for $N_{\rm use}=10$, respectively.  Their different nonlinear dependence on the same estimated visibility leads to different sampling responses.  In panel (a), inference with the collective model approaches the $c=1$ resource curve even when the true environment has $c=0.65$.  Panel (b) compares the mean absolute error over 600 records at $t=6$ as $N_s$ increases from $100$ to $5000$.  With the correct model, the error decreases with sample size.  With an incorrect correlation model, it approaches the systematic bias $|R_{{\rm RE},{\rm com}}^{(0.5)}(6)-R_{{\rm RE},c=0.65}^{(0.5)}(6)|\simeq0.279$.

The displayed bands describe repeated-sampling distributions under the specified model, rather than confidence intervals inferred from experimental records.  These comparisons separate finite-sample fluctuations from correlation-model bias and identify the calibration information needed to infer resource reliability from measurement counts.

\section{Conclusion and outlook}
\label{sec:conclusion}

The central result of this work is a resource-based route to quantum-device reliability when loss of the specified function cannot be represented by occupation of a fixed failure subspace.  By linking an established resource measure to an operational task and a stated ideal capacity, the remaining resource defines a quantitative reliability benchmark.  This shifts the assessment from subspace occupation to the resource still available for the task.  The construction complements projector-based reliability where the latter is applicable, but does not identify a normalized resource with a universal classical survival probability.

The QRNG and interferometer examples give this proposal concrete operational content.  For the trusted QRNG reference architecture, relative-entropy coherence determines the optimal asymptotic coherent-qubit yield.  The corresponding target-slot probability approaches resource reliability only in the specified vanishing-error limit.  A finite-block yield and its output-quality guarantee remain separate quantities.  For the ideal two-port interferometer, the logarithmic coherence determines the calibrated correct-port probability, and independent complete uses give the associated all-correct probability.  These task-level event probabilities explain how resource reliability can inform a probabilistic assessment without being a first-failure-time distribution.

The spin-boson results show why the resource and task specifications matter.  At the same isolated-spin attenuation, collective and spatially resolving baths can leave different many-spin reliabilities.  Collective noise preserves within-sector coherence for product inputs, whereas GHZ branch coherence decays with the enhanced collective exponent.  Retained basis coherence need not preserve response to a given phase encoding, as the full-phase readout illustrates.  These comparisons concern the stated pure-dephasing model and do not imply uniform protection of every task by environmental correlations.

Finite-record assessment makes the proposal experimentally testable under explicit calibration assumptions.  The interferometer estimator follows from a constrained binomial likelihood.  The QRNG model-assisted estimator additionally requires validation of the many-spin noise correlations.  Our simulations distinguish sampling fluctuations from model-mismatch bias and show that increasing the number of probe measurements cannot remove the latter.  Their sampling bands are not experimental confidence intervals or device-independent guarantees.  Applying the framework to an implemented device requires incorporating control errors, loss, drift, and inter-use memory.  A secure QRNG rate would further require finite-block analysis, treatment of information available to an adversary, and randomness extraction.

The present construction concerns tasks tied to a single coherence resource.  Extending it to devices that require several resources will require a task-dependent rule for their joint use, consistent with their conversion constraints and trade-offs~\cite{sparaciari2020firstlaw}.  Such an extension should retain the principle established here: when a failure subspace does not describe functional degradation, reliability assessment can instead be based on the operational resources that the device still supplies.

\begin{acknowledgments}
This work was supported by the Science Challenge Project (Grant No.~TZ2025017) and the National Key Research and Development Program of China (Grants No.~2021YFA1402104 and No.~2021YFA0718302).
\end{acknowledgments}

\bibliographystyle{apsrev4-2}
\bibliography{quantum_resource_reliability}

\end{document}